\documentclass[fleqn,usenatbib]{mnras}

\usepackage{newtxtext,newtxmath}

\usepackage[T1]{fontenc}
\usepackage{ae,aecompl}

\usepackage[]{graphicx}	
\usepackage{amsmath}	
\usepackage{amssymb}	
\usepackage{gensymb}

\title[Mass transport limits black hole growth]{Mass transport in galaxy discs limits black hole growth to sub-Eddington rates}

\author[Eastwood, Khochfar and Trew]{
Daniel S. Eastwood$^{1}$\thanks{E-mail: deastw@roe.ac.uk (DSE)}
Sadegh Khochfar$^{1}$
and Arthur Trew$^{2}$
\\
$^{1}$Institute for Astronomy, University of Edinburgh, Royal Observatory, Edinburgh EH9 3HJ, UK\\
$^{2}$School of Physics and Astronomy, University of Edinburgh, Edinburgh, EH9 3JZ, UK
}

\date{Accepted XXX. Received YYY; in original form ZZZ}

\pubyear{2019}

\begin{document}
\label{firstpage}
\pagerange{\pageref{firstpage}--\pageref{lastpage}}
\maketitle

\begin{abstract}
Super-massive black holes (SMBHs) observed to have masses of $M_\bullet \sim 10^9 \, \mathrm{M_\odot}$ at $z\gtrsim6$, $<1$ Gyr after the Big Bang, are thought to have been seeded by massive black holes which formed before growing concurrently with the formation of their host galaxies. We model analytically the idealised growth of seed black holes, fed through gas inflow from growing proto-galaxy discs. The inflow depends on the disc gravitational stability and thus varies with black hole and disc mass. We find that for a typical host halo, the efficiency of angular momentum transport, as parametrised by the disc viscosity, is the limiting factor in determining the inflow rate and the black hole accretion rate. For our fiducial case we find an upper black hole mass estimate of $M_\bullet \sim 1.8 \times 10^7 \, \mathrm{M_{\odot}}$ at $z=6$. Only in the extreme case of $\sim 10^{16}$ M$_{\odot}$ haloes at $z=6$ produces SMBH masses of $\sim 10^9$ M$_{\odot}$. However, the number density of such haloes is many orders of magnitude below the estimated 1 Gpc$^{-3}$ of SMBHs at $z=6$, indicating that viscosity driven accretion is too inefficient to feed the growth of seeds into $M_\bullet \sim 10^9 \, \mathrm{M_\odot}$ SMBHs by $z \sim 6$. We demonstrate that major mergers are capable of resolving the apparent discrepancy in black hole mass at $z=6$, with some dependence on the exact choice of orbital parameters of the merger.
\end{abstract}



\section{Introduction}

Super-massive black holes (SMBHs) have been observed to power quasars at redshifts as high as $z\gtrsim7$ \citep{Fan2006a, Mortlock2011, Banados2018}. In $\Lambda$CDM, the Universe is understood to be only $\sim800$ Myr old at this epoch \citep{Planck2016} and yet these SMBHs with masses of $M_\bullet \sim10^9\;\mathrm{M_\odot}$ are abundant with comoving number densities of around $\phi \sim 1 \, \mathrm{Gpc^{-3}}$ \citep{Fan2003}. If the remnants of the first generation of stars are the progenitors of SMBHs, they must have grown roughly seven orders of magnitude over extremely short timescales. The growth rates necessary for this are above the classical upper limit on the accretion of material, the Eddington limit \citep[see however,][]{Inayoshi2016,Pacucci2017}.

Studies have looked to resolve this issue by introducing alternative black hole formation mechanisms with the capability of generating higher mass progenitors for SMBHs, known as massive seed black holes \citep[see][for a review]{Volonteri2010}. A higher mass seed black hole has the advantage of requiring a lower average specific growth rate to become super-massive, given the same period in which to grow. One such alternative formation mechanism is the direct collapse of primordial gas in a $T_{\rm vir}\gtrsim10^4$ K halo, prior to the formation of a galaxy, creating a single massive seed with $M_\bullet \sim 10^{4-6} \; \mathrm{M_\odot}$ \citep{Bromm2003, Agarwal2012b, Agarwal2014a}. Known as direct collapse black holes (DCBHs), these objects are thought to form in metal-free haloes \citep{Begelman2006} with a local intensity of Lyman-Werner (LW) photons\footnote{UV photons with energies of $11.2 - 13.6$ eV capable of dissociating $\mathrm{H}_2$ molecules, though hereafter also used to loosely refer to photons with energies $> 0.76$ eV, capable of causing the photo-detachment of $\mathrm{H}^{-}$.} high enough to photo-dissociate any $\mathrm{H}_2$ molecules, along with a sufficient intensity of photons to photodetach any $\mathrm{H}^{-}$ \citep{Shang2010, Agarwal2016}. In the absence of $\mathrm{H}_2$ and $\mathrm{H}^{-}$, the primordial gas cannot cool below the atomic cooling limit at $T_{\rm gas} \sim 8000$ K, avoiding fragmentation during the subsequent gravitational collapse \citep{Spaans2006, Dijkstra2008, Shang2010}.

Even with the most optimistic seed masses, maintaining the necessary growth rate for seed black holes to become super-massive ($M_\bullet \sim 10^9\;\mathrm{M_\odot}$) is still a challenge \citep{Dubois2015, Latif2018}. Simulations of massive black holes traditionally use Bondi-Hoyle-Lyttleton accretion (\citealt{Bondi1944}, \citealt{Bondi1952}, \citealt{Edgar2004}, see however, \citealt{Hopkins2011}, \citealt{AnglesAlcazar2015} and \citealt{Beckmann2018}), where the accretion rate is related to the properties of the local medium and the black hole mass, to calculate the growth rate of the black hole. Theoretically, this could result in super-Eddington growth rates. Indeed, there are indications that super-Eddington accretion could be viable through a number of scenarios \citep{Ohsuga2005, Pacucci2015, Inayoshi2016, Pacucci2017, Takeo2019arXiv}. However, high-resolution numerical simulations of massive black holes that account for the feedback from AGN and stars struggle to reach the Eddington accretion rate \citep{Johnson2011, Dubois2015, Latif2018}.

High intensities of LW photons are found in close proximity to massive galaxies with significant, early star formation, making  haloes in the vicinity of the first galaxies ideal formation sites for DCBHs \citep{Dijkstra2008, Agarwal2018}. Indeed, cosmological simulations have shown that DCBH formation can occur throughout the Universe at $z\sim10-20$, with such haloes particularly favoured as formation sites \citep{Agarwal2012b, Agarwal2014a, Wise2019}. The growth histories of haloes within these simulations show that once a DCBH is formed, the subsequent merger of the DCBH host halo with a more massive central galaxy's halo is likely. This means that DCBH seeds are unlikely to remain isolated throughout their evolution and at least one major merger should occur following their formation \citep{Dijkstra2008}.

Merger events are known to drive strong starbursts in galaxies and are capable of feeding black hole growth \citep{Hopkins2006, Li2007, DiMatteo2008}. Furthermore, there is evidence that the galaxy-black hole scaling relations observed at lower redshifts \citep[see, e.g.][]{Kormendy2013, Heckman2014} are likely reached through growth driven by mergers \citep{DiMatteo2005, Schawinski2006, Hirschmann2010, Lamastra2010}. In this scenario, mergers are thought to instigate bursts of star formation \citep{Mihos1996} and drive strong inflows of gas to feed quasar activity \citep{Springel2005b}. The resulting energy released by an active quasar heats and expels gas, quenching star formation while potentially enhancing further black hole growth \citep{DiMatteo2005, Hopkins2006}. However, the significance of mergers in driving the growth of SMBHs has more recently been called into question \citep[see, e.g.][]{Fanidakis2012, Hirschmann2012, DelMoro2016, Villforth2017, Hewlett2017, Steinborn2018}. Observations have found no statistical evidence for an enhanced merger fraction in galaxies with AGN, and indicate the role of mergers in driving activity is, at best, secondary to other fuelling mechanisms \citep[see, e.g.][]{Villforth2017}. Though, merger-driven activity is at least slightly more significant at higher redshift \citep{DelMoro2016, Hewlett2017}. These observational results are somewhat controversial due to AGN survey limitations \citep[e.g.][]{Juneau2013} and the challenges of observing evidence of merger-driven activity as a result of the long time delay between merger events and the subsequent AGN activity \citep{Schawinski2010}. However, recent simulations also find mergers are insignificant for driving black hole growth, except for $z\gtrsim2$ where the expected higher merger rates come into play \citep{Steinborn2018}. With merger-driven black hole growth potentially more significant at higher redshift, the role of mergers in growing $z\sim6$ quasars cannot be ignored.

The question of SMBH growth is further complicated when considering larger scale effects, such as the growth of host halo itself and the potential co-evolution of the black holes with their host galaxies. Within halos where DCBHs are predicted to form, the conservation of angular momentum will result in the formation of proto-galactic discs \citep{Oh2002}. These can then be fed by the accretion onto the host halo from cold gas streams \citep{Dekel2009}. Gas that is not depleted through star formation will still have to lose angular momentum efficiently to feed any black hole growth. As already stated, merger events can drive gas inflow through the efficient transport of angular momentum \citep{Barnes1996, DOnghia2006, Hopkins2009}. Processes internal to the disc can also result in gas inflow. Given the right circumstances, gravitational instabilities are thought to be capable of efficiently driving the inflow of material in self-gravitating discs \citep{Toomre1964, Rice2005}. Indeed, different models of galactic discs have shown gravitational instabilities are capable of providing inflow which supports black hole growth \citep{Lodato2006, Devecchi2010, Hopkins2011, AnglesAlcazar2015}.

In this study we model the growth of a black hole and its host galaxy from the formation redshift of the seed black hole to $z\sim6$. Building on the model discussed in \citet{Eastwood2018a} we investigate how the growth of seed black holes can be limited by the inflow of gas through a growing viscous (proto-)galactic disc. We model the inflow rate from a viscous disc and show how the flow is a function of overall gravitational stability of the disc. We neglect energetic feedback effects and focus on predicting upper limits to the possible feeding of black holes through viscosity driven gas inflow. We start in Section~\ref{sect:method} by introducing the model and describing how the inflow rate of gas from the disc onto the black hole is estimated using the two extreme situations of a conservative and optimistic inflow rate in the viscous disc. Section~\ref{sect:results} describes the consequences of the inflow rate on the growth of the black hole and investigates how the final mass of the black hole depends on the model parameters. In Section~\ref{sect:mergers} we demonstrate how mergers can resolve the otherwise apparent discrepancy between observations and our upper estimate in the black hole mass at $z=6$. Finally, we summarise our findings and discuss the implications for the sites of massive seed formation and the evolution of SMBHs hosts in Section~\ref{sect:summary}.

\section{Methodology} \label{sect:method}

The methodology presented here builds on the model discussed in \citet{Eastwood2018a} which focuses on the growth of a proto-galaxy in an atomic cooling halo, modelled as an isothermal sphere, that is free of molecular hydrogen due to a high intensity LW-background and with a DCBH at the centre. Here we discuss briefly that model and outline the changes made in undertaking this study.

\subsection{Halo growth}

The model initial conditions place a massive black hole ($M_\bullet= 10^{4-6}\, \mathrm{M}_\odot$) at the centre of a proto-galactic disc at some formation redshift, $z_{\rm i} \sim 20 - 10$, within a $T_{\rm vir} = 8000$ K dark matter halo. The disc mass is calculated at any point from the baryon mass of the halo once the black hole mass is subtracted, assuming the universal baryon fraction $f_{\rm b}=0.17$ in the halo ($M_{\rm d}(t)= f_{\rm b}\, M_{\rm h}(t) - M_\bullet(t)$). The disc is assumed to be isothermal with a gas temperature of $T_{\rm g}=8000$ K, consistent with the absence of molecular cooling in primordial gas. The system is then allowed to grow with the halo and disc growth being fed through cosmological accretion with halo growth rate parametrised by $\zeta$. 

\begin{equation}
    \frac{d M_{\rm tot}}{dz} = - \zeta M_{\rm tot}
    \label{halogrowth}
\end{equation}

\noindent The mean halo growth rate from $\Lambda$CDM provides our fiducial value of $ \langle \zeta \rangle=0.806$  \citep{Neistein2008}. Using the scatter in the growth rate of dark matter halos from \citet{Genel2008}, the standard deviation in $\zeta$ can be calculated as $\sigma_\zeta= \langle \zeta \rangle(2.5/(1+z))^{0.2}=0.656$ at $z=6$ by assuming the growth rate is a linear function of the halo mass. $\langle \zeta \rangle + \sigma_\zeta = 1.462$ is taken as an extreme upper bound. A further constraint is placed on the maximum growth rate of our model haloes from the known number density of massive quasars at $z \sim 6$ of around $\phi \sim1 \, \mathrm{Gpc^{-3}}$ \citep{Fan2003}. This number density corresponds to haloes with masses $M_{\rm tot} \ge M_{\rm tot, \,lim} \approx 10^{13} \, \mathrm{M_\odot}$ \citep{Sheth2001, Murray2013}. Thus we can calculate for a given $z_{\rm i}$ (and corresponding atomic cooling halo mass) the halo growth rate, $\zeta_{\rm max}\left(z_{\rm i}\right)$, which will result in $M_{\rm tot}=M_{\rm tot, \,lim}$. At higher halo growth rates with $\zeta>\zeta_{\rm max}\left(z_{\rm i}\right)$, the halo mass at $z=6$ will be higher $M_{\rm tot}>M_{\rm tot, \,lim}$ and the abundance of such haloes would be lower than the abundance of SMBHs. As we are only considering the seeding of SMBHs, $\zeta_{\rm max}$ provides a further upper constraint to the halo growth rate for earlier formation times ($z_{\rm i} \sim 20-15$). Note the direct mapping used between $\zeta$ and the total halo mass $M_{\rm tot}(z=6)$ for a given $z_{\rm i}$.

\citet{Agarwal2016a} modelled the growth history of DCBH host candidate, CR7 \citep{Sobral2015,Bowler2017} and found mass assembly histories for the each of the subhaloes in the model. By fitting an exponential to the median subhalo mass at each whole number redshift, we find a halo growth rate parameter of $\zeta_{\rm CR7}=0.568$ which we take as our lower bound.


\subsection{Disc instability and inflow rates}

\begin{figure}
	\includegraphics[width=\columnwidth]{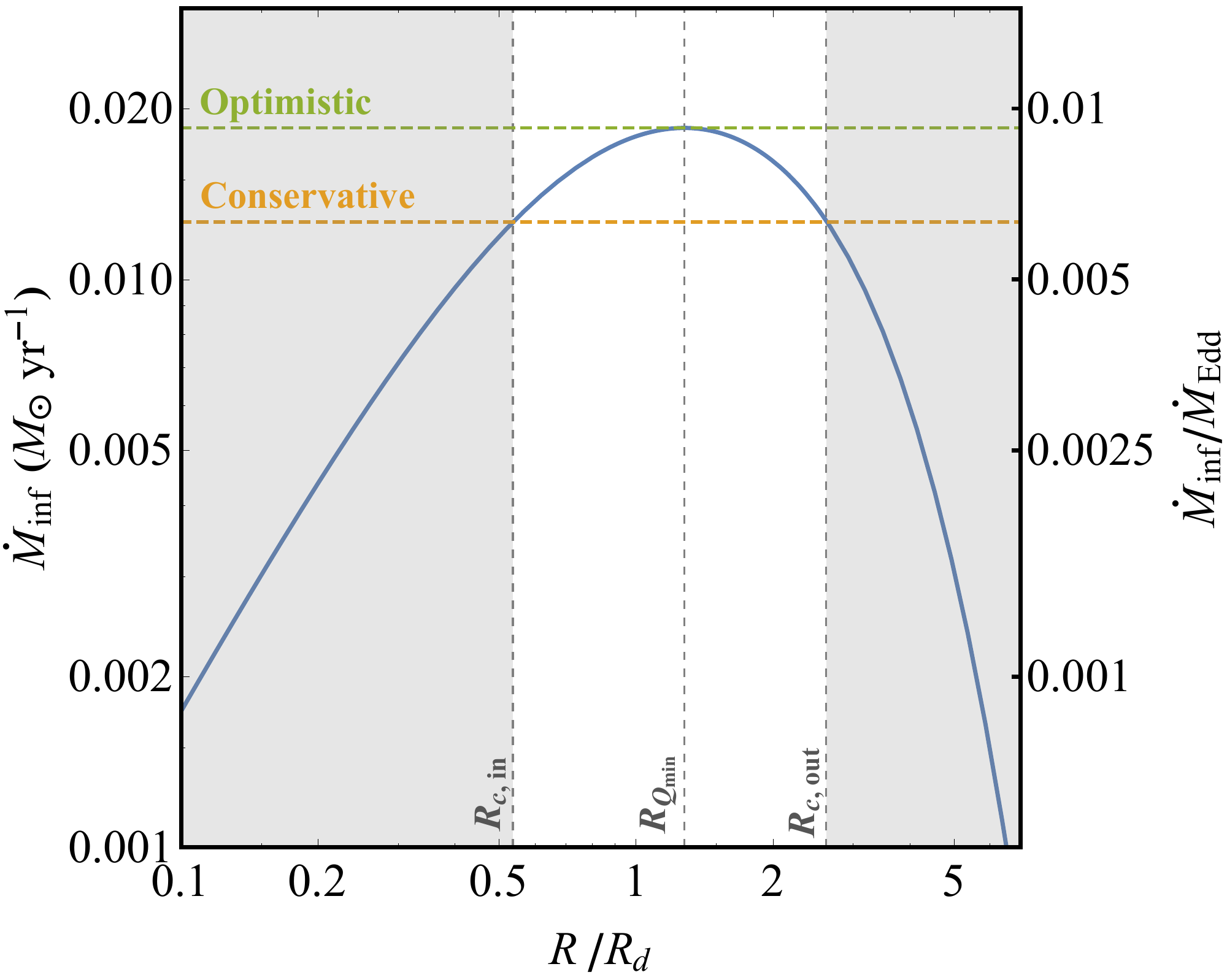}
    \caption{The inflow rate of gas as a function of radius in a $M_{\rm d}\sim10^8 \, \mathrm{M}_\odot$, gaseous disc centred on a $M_\bullet=10^8 \, \mathrm{M}_\odot$ black hole. The radius is shown as a fraction of the disc scale radius ($R_{\rm d} = 40$ pc). The greyed out regions show the portions of the disc where $Q>1$. The blue line shows the inflow rate while the horizontal, green and orange dashed lines show the optimistic and conservative inflow rates respectively. The vertical dashed lines from left to right indicate the radii where $Q=1$, $Q=Q_{\rm min}$, and again $Q=1$. The right hand vertical axis shows the inflow rate as a fraction of the Eddington growth rate onto the black hole.}
    \label{fig:mdotinfR}
\end{figure}

Gravitational instabilities in discs can drive angular momentum transport and the inflow of material \citep{Toomre1964}. Disc instability is often parametrised using the Toomre-$Q$ parameter.

\begin{equation}
    Q=\frac{\sigma \kappa}{\pi G \Sigma}
    \label{Toomre}
\end{equation}

where $\sigma$ is the velocity dispersion, which we assume in this study to be dominated by the gas sound speed, $c_{\rm s}$, and $\Sigma$ is the total surface density given by the radial profile. For simplicity, the gas and stellar components are assumed to follow the same profile, $\Sigma(R)=\Sigma_0 \exp{-R/R_{\rm d}}$. As we model the dark matter halo profile as an isothermal sphere, the disc scale radius, $R_{\rm d}$, scales with the virial radius of the halo and the spin parameter, $\lambda$ \citep{Mo1998}. We assume throughout this paper $\lambda = 0.05$, corresponding to the mean \citep{Mo1998}. The epicyclic frequency, $\kappa$, measures the differential rotation and is a function of the radial profile of the angular frequency, $\Omega(R)$. It can be expressed as the following:

\begin{equation}
    \kappa = 2 \Omega \sqrt{1 + \frac{1}{2}\frac{d \ln{\Omega}}{d \ln{R}}}
    \label{kappa}
\end{equation}

The star formation rate can be calculated using the same method as in the previous study \citep{Eastwood2018a} in the region of the disc where $Q<1$. However, rather than the stellar mass remaining where it forms in the disc, the stellar surface density is assumed to follow the exponential radial profile at all times. This does not significantly impact the total stellar mass of the disc however it will change how the stability of the disc varies with radius.

Generally, mass transport in discs is described using some viscosity, $\nu=\alpha_\nu c_{\rm s} H$, where $c_{\rm s}$ is the disc sound speed in the mid-plane, $H\sim c_{\rm s}/\Omega$ is the disc scale height, and $\alpha_\nu$ is the viscosity parameter. $\alpha_\nu$ acts as an efficiency parameter which must account for the physical mechanism behind the viscosity. The viscosity parameter has been determined through observations of discs in various physical scenarios \citep{King2007} with a range of $0.1\leq \alpha_\nu \leq 0.4$. However, there is evidence to suggest inflow with $\alpha_\nu \gtrsim 0.06$ should result in the formation of clumps and the disruption of the transport of mass inwards \citep{Rice2005, Lodato2006}. Previously, studies of Mestel discs ($\Sigma(R)=\Sigma_0$) have related $\alpha_\nu$ to the global Toomre-$Q$ parameter \citep{Lodato2007, Devecchi2009, Devecchi2010}. Here, an exponential profile is centred on our point mass black hole, both resulting in $\kappa$ and $Q$ being functions of $R$. The relationship between a global $\alpha_\nu$ and $Q(R)$ in this setting becomes unclear as $Q$ is now a local quantity. For our purposes, we investigate a range of $\alpha_\nu = 0.06 - 0.4$, bearing in mind that a decrease in $\alpha_\nu$ will only act to decrease the inflow rate.

The inflow rate of gas in the disc as a result of this viscosity is defined as,

\begin{equation}
    \dot{M}_{\rm inf} = \frac{\alpha_\nu c_{\rm s}^2 \pi \Sigma_{\rm g}}{2 \Omega} \left| \frac{d \ln{\Omega}}{d \ln{R}} \right|
    \label{mdotinf}
\end{equation}

\noindent where $\Sigma_{\rm g}$ is the gas surface density. For a gaseous disc, this can then be rearranged to be a function of the Toomre-$Q$ parameter:

\begin{equation}
    \dot{M}_{\rm inf} = \frac{\alpha_\nu c_{\rm s}^3}{G} \frac{1}{Q_{\rm g}} \left| \frac{d \ln{\Omega}}{d \ln{R}} \right| \sqrt{1 + \frac{1}{2}\frac{d \ln{\Omega}}{d \ln{R}}}
    \label{mdotinf2}
\end{equation}

As the inflow rate described by equations~\ref{mdotinf} and~\ref{mdotinf2} has a radial dependence (see Figure~\ref{fig:mdotinfR}), to avoid calculating the full evolution of the disc profile we must make an estimate for the inflow rate from the disc to centre of the galaxy. Though the gradient of the angular frequency ($d \ln{\Omega}/d \ln{R}$) does vary with radius and $M_\bullet/M_{\rm d}$, it remains $\sim \mathcal{O}(-1)$ with $-2/3 \gtrsim d \ln{\Omega}/d \ln{R} \gtrsim -3/2$ for our setup. The largest contribution to the radial variation therefore comes from the profile of $Q_{\rm g}$. This leads to the inflow rate peaking at a radius roughly equivalent to the radius at which the stability parameter is minimised, i.e. where the disc is most unstable. If the effective inflow rate of the disc is determined to be the value of $\dot{M}_{\rm inf}$ where $Q$ is minimised with respect to $R$, this is roughly an upper limit on the rate at which a central gas reservoir can be fed from accretion in the galaxy disc. We take $\dot{M}_{\rm inf}(R_{Q_{\rm min}})$ to be an estimate on the upper limit on the inflow rate and therefore refer to it as the optimistic inflow rate. This is indeed an upper limit estimate as a full calculation of the profile where Equation~\ref{mdotinf2} is used to calculate $\dot{M}_{\rm inf}(R)$ would result in a flattening of the surface density profile. The lower surface density in the unstable region (as the mass flows inwards), will raise $Q$ and lower the inflow rate. Of course, deviations in the radial profile of the disc away from an exponential would change the $Q$ profile beyond the unstable region. However, following the full calculation of the surface density profile would only result in a higher total inflow of gas than the optimistic estimate if the disc growth from cosmic accretion no longer followed the assumed exponential profile. Such a scenario is beyond the scope of this paper and we defer more realistic investigations to a subsequent study using cosmological simulations.

Inflow is calculated within the region where the disc is unstable to the gravitational instabilities which are driving the viscosity (i.e. for R where $Q<1$). For gas to reach the centre of the disc, the inflow needs to continue down to $R<<R_{\rm d}$. As we do not calculate the full evolution of the disc profile here, we assume that inflow is able to continue at smaller radii by maintaining an equilibrium of sorts, whereby the inflow rate at the inner $Q=1$ radius is balanced by the inflow rate into the gravitational sphere of influence of the black hole. We therefore use the inflow rate at the inner radius where $Q=1$, $R_{\rm c, in}$, as our conservative inflow rate. This is a conservative lower limit estimate as only lower inflow rates would be determined if $\dot{M}_{\rm inf}(R \sim R_\bullet)$ was used (though $R=R_\bullet$ is outside of the $Q<1$ region). Note this does not influence the main findings of this paper for which we focus more on the optimistic inflow rate. The optimistic and conservative estimates of the inflow rate are shown on Figure~\ref{fig:mdotinfR} by the green and orange dashed lines respectively.

\subsection{Black hole growth}

\begin{table}
	\centering
	\caption{Table of model parameters varied in this study.}
	\label{tab:para_table}
	\begin{tabular}[h]{l|l|l} 
		\hline
		Parameter & Definition & Fiducial (Range)\\
		\hline
		$\alpha_\nu$ & viscosity parameter & 0.06 ($0.06 - 0.4$)\\
		$M_{\rm seed}$ & black hole seed mass & $10^{6}$ ($10^{4 - 6}$) $\mathrm{M_{\odot}}$\\
		$z_{\rm i}$ & seed formation redshift & 20.0 ($20.0 - 10.0$)\\
		$\zeta$ & halo growth rate & 0.806 (0.586, 0.806, 1.462) \footnotemark\\
	\end{tabular}
\end{table}
\footnotetext{A further constraint on the growth rate comes from requiring the halo mass at $z=6$ corresponds to a number density higher than $1 \, \mathrm{Gpc^{-3}}$ on the halo mass function. This constraint depends on $z_{\rm i}$.}

The black hole is fed by gas from a reservoir with mass, $M_{\rm res}$, which is itself fed from the inflow of the disc. The mass of the reservoir is thus determined by the balance of the inflow rate and the accretion rate of the black hole, $\dot{M}_\bullet$, giving:

\begin{equation}
    \dot{M}_{\rm res} = \dot{M}_{\rm inf} - \dot{M}_{\bullet}
    \label{dotmres}
\end{equation}

The reservoir is assumed to be under the direct gravitational influence of the black hole. From this we calculate the accretion rate of the black hole from the mass available in the gas reservoir via,

\begin{equation}
    \dot{M}_\bullet = \mathrm{Min}\left[\frac{M_{\rm res}}{t_{\rm acc}} = \varepsilon \frac{M_{\rm res}}{t_{\rm dyn}}, \;\dot{M}_{\rm Edd}\right]
    \label{dotmbh}
\end{equation}

\noindent where $\dot{M}_{\rm Edd} = M_\bullet/t_{\rm Sal}$ is the Eddington limit accretion rate, $t_{\rm Sal}\sim 0.05$ Gyr is the Salpeter timescale with 10\% radiative efficiency \citep{King2008}, and $t_{\rm acc} = t_{\rm dyn}/\varepsilon $ is the accretion timescale with the efficiency parameter $\varepsilon<1$. The value of $\varepsilon$ is largely unknown but we note that generally it does not determine the growth history of the black hole which is largely controlled by the inflow rate feeding the reservoir (see below). The timescale is determined at the radius of the sphere of influence of the black hole, $R_\bullet$, i.e. $t_{\rm dyn} = 1/\Omega\left(R_\bullet\right)$, where $\Omega$ is the angular velocity. Inside this radius the mass of the black hole will dominate the potential and any material will eventually be accreted onto the black hole subject to the loss of angular momentum\footnote{$R_\bullet$ does not significantly affect our findings as $R_{\rm c, in} >> R_\bullet$ \citep{Eastwood2018a}, and thus the inflow that feeds the reservoir, which primarily determines the evolution of the black hole (see Section~\ref{sect:results}), will occur over longer timescales and is independent of $R_\bullet$.}. $R_\bullet$ is calculated as the radius at which the gravitational potential due to the black hole balances that of the halo and the thermal energy of the disc,

\begin{equation}
    R_\bullet = \frac{G M_\bullet}{c_{\rm s}^2 + V_{C,\,h}^2}
    \label{rbh}
\end{equation}

\noindent where $c_{\rm s}$ is the sound speed of the gas in the disc and $V_{C,\,h}$ is the circular velocity due to the gravitational potential of the halo.

If no limit is imposed on the growth of the black hole, such as Eddington limit in Equation~\ref{dotmbh}, we can take advantage of the short accretion timescale to make a simplification to the model. The black hole is able to grow via an accretion disc which exists within the sphere of influence of the black hole. The scale of the accretion disc is therefore significantly smaller than galactic disc scales, and the inflow of material from gravitational instabilities in the galactic disc are not directly feeding the black hole. However, to stringently test the capability of viscosity driven inflow to feed the growth of massive black holes, we can assume that the inflow from the disc is able to directly feed the black hole (i.e. $\dot{M}_\bullet = \dot{M}_{\rm inf}$). With this simplification, we find an upper limit on black hole growth fed via viscosity driven inflows. As the accretion timescale is much shorter than the infall timescale, using the simplification $\dot{M}_\bullet = \dot{M}_{\rm inf}$ will not significantly impact the growth of the black hole in the model.

We stress this simplification cannot be made where the accretion timescale from the reservoir would be longer than the inflow timescale, such as when limiting the growth by the Eddington rate or for very small values of $\varepsilon$. As we are focusing on calculating an upper limit for black hole growth this does not affect our main findings, however, whenever the Eddington limit is imposed in a model, the reservoir is included in the calculation.

\section{Results} \label{sect:results}

\begin{figure*}
\centering

  \begin{tabular}{@{}c@{}}
    \includegraphics[width=\textwidth]{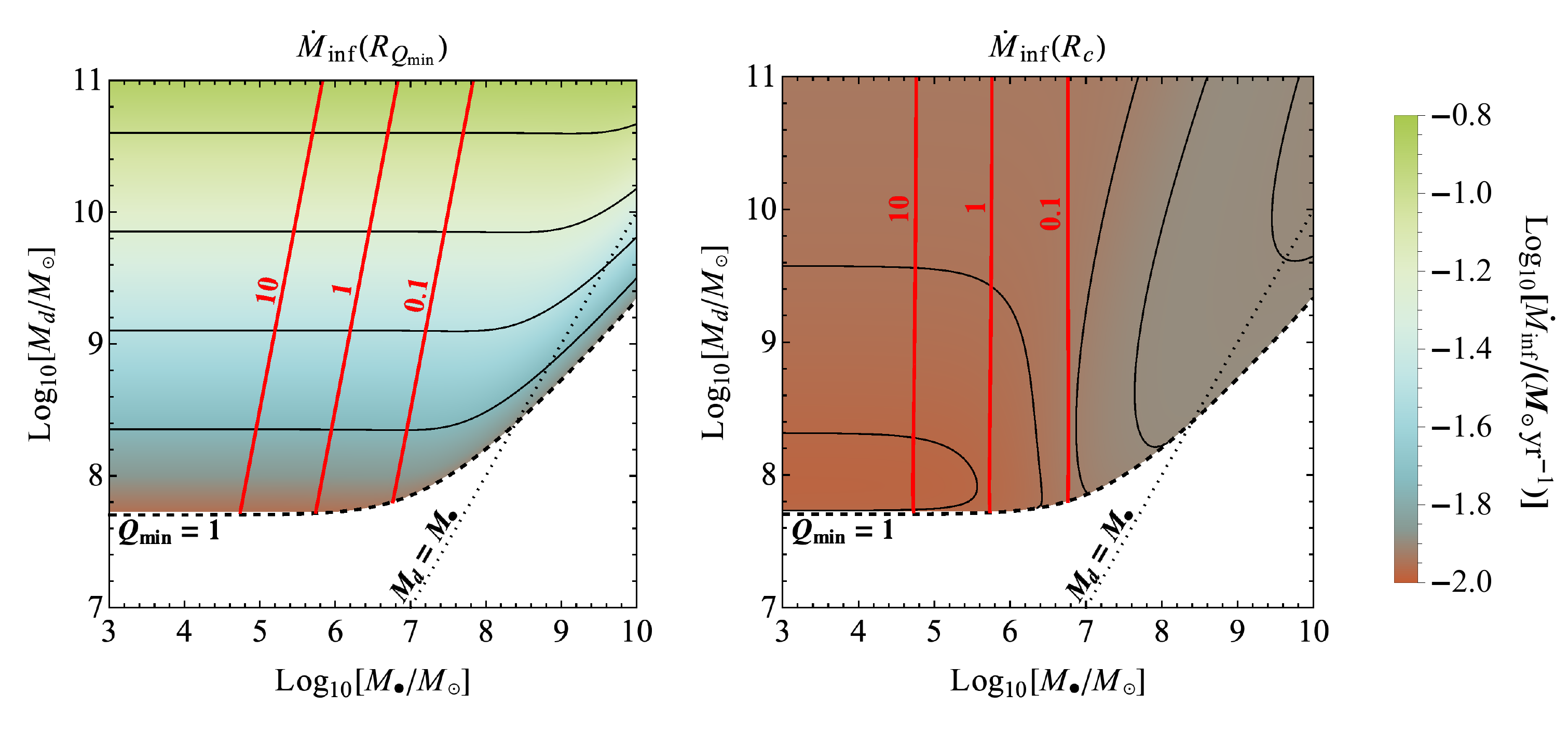}\\
    \includegraphics[width=\textwidth]{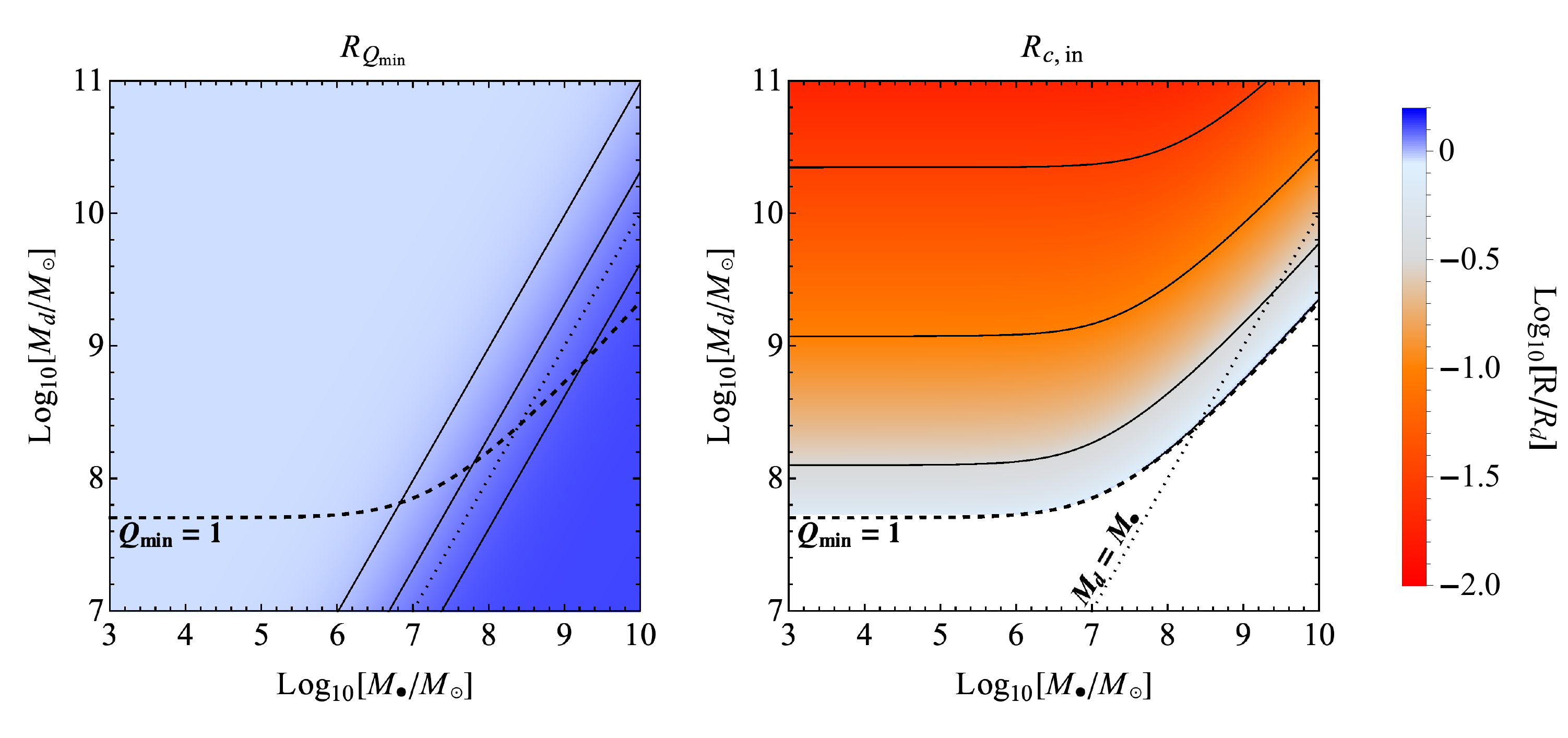}
  \end{tabular}
    \caption{As a function of disc and black hole mass each panel shows the following: the top row shows the inflow rate of gas from the disc onto the reservoir of gas which feeds the black hole (the red lines indicate different values of $\dot{M}_{\rm inf}/\dot{M}_{\rm Edd}$ as indicated by the red labels, i.e. the $\dot{M}_{\rm inf}/\dot{M}_{\rm Edd}=1$ line shows where the inflow rate goes from super-Eddington to sub-Eddington from left to right). The top left panel shows the optimistic inflow rate i.e. the rate where the disc is most unstable (at $R = R_{Q_{\rm min}}$). The top right panel shows the conservative inflow rate i.e. the rate where the disc is only marginally unstable (at $R = R_{Q = 1}$). The bottom left and right show the values for these radii. In each case the dark matter halo mass scales with the baryon mass ($M_{\rm b}=M_{\rm d} + M_{\bullet}$) so that $f_{\rm b}=0.17$ and $R_{\rm d}$ scales with the virial radius of the halo \citep{Mo1998}. The inflow rate is determined using the fiducial value for the viscosity parameter $\alpha_\nu=0.06$ (note this is just a constant factor in $\dot{M}_{\rm inf}$). The thick dashed line shows where $Q_{\rm min} = 1$, i.e. above this line the disc is partly unstable and inflow can occur. Note, that the slight dependence of the conservative accretion rate on the black hole mass is a result of the small variation in $d \ln{\Omega}/d \ln{R}$. The dotted line on each panel indicates the $M_{\rm d}=M_{\bullet}$ line.}
    \label{fig:mdotinf_mdmbh}
\end{figure*}

\begin{figure*}
\centering
\begin{tabular}{lr}
  \includegraphics[width=\columnwidth]{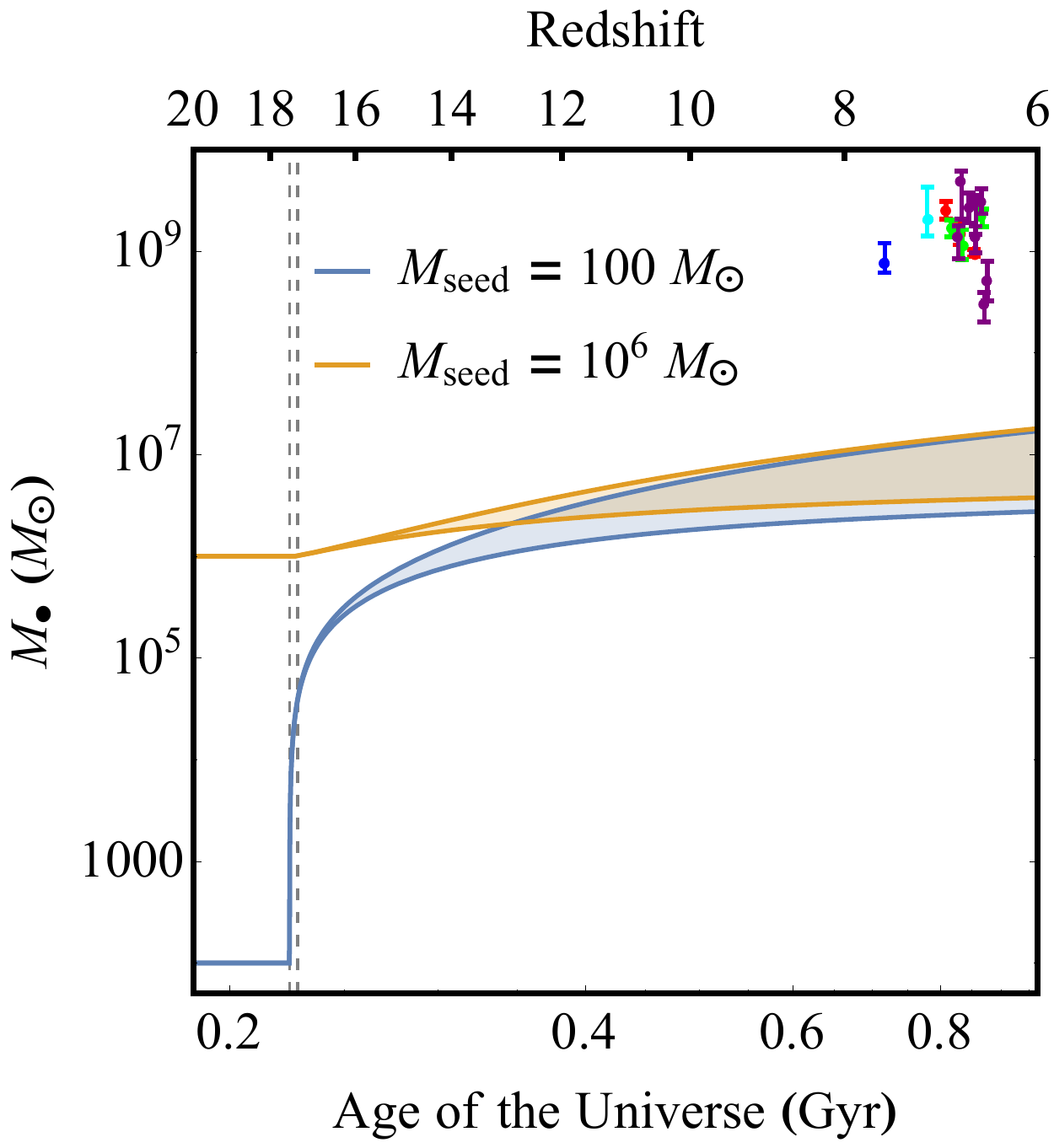}&
  \includegraphics[width=\columnwidth]{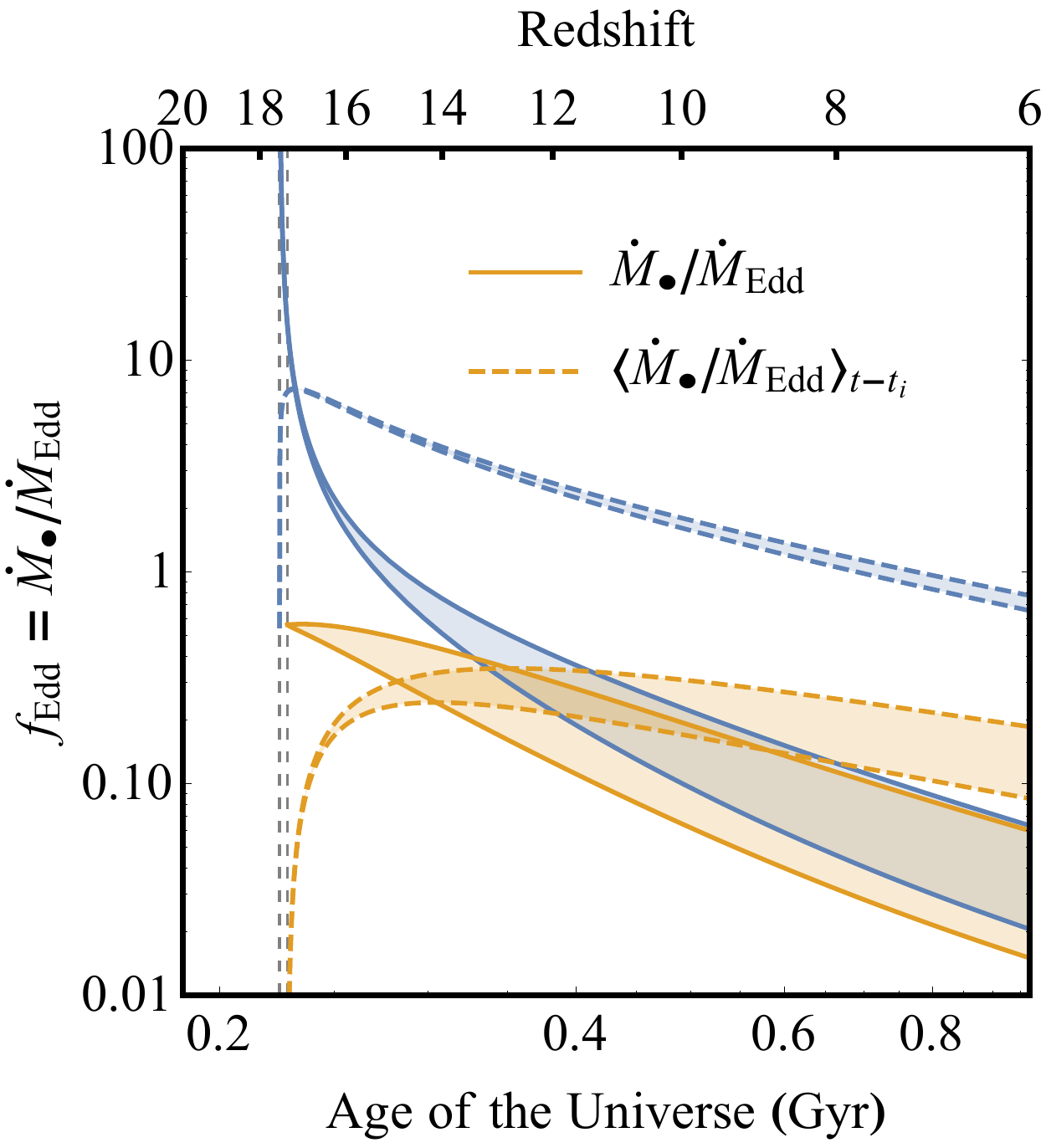}
\end{tabular}
    \caption{The growth history of the black hole and the gas reservoir which feeds the black hole for the fiducial model with seeds of initial masses $M_{\rm seed}=10^6 \, \mathrm{M_\odot}$ and $M_{\rm seed}=100 \, \mathrm{M_\odot}$, which form at $z_{\rm i}=20$. The other model parameters are set to their fiducial (see Table~\ref{tab:para_table}) and the black hole accretion rate was not capped by the Eddington limit. The possible growth histories are shown as the shaded regions between the curves calculated using $\dot{M}_{\rm inf}(R_{Q_{\rm min}})$ and   $\dot{M}_{\rm inf}(R_{\rm c,\, in})$ as the upper and lower estimates for the inflow rate onto the gas reservoir respectively. Each panel shows the following: Left: The evolution of the mass of the black hole. The coloured data points are from observations of high-redshift quasars (cyan: \citet{Mortlock2011}, red: \citet{DeRosa2014}, purple: \citet{Mazzucchelli2017}, blue: \citet{Banados2018}, green: \citet{Reed2019arXiv}). The vertical dotted lines indicate the times at which the disc first becomes unstable ($Q_{\rm min}<1$). In the lower seed mass case this is reached marginally earlier. Right: The evolution of the growth rate onto the black hole as a fraction of the Eddington limit. The Eddington fraction averaged over the time since formation, $t -t_{\rm i}$ (see Equation~\ref{eddave}), is also shown using the regions bound by the dashed curves.}
    \label{fig:mevo_mseed}
\end{figure*}

\begin{figure*}
\centering
\begin{tabular}{lr}
  \includegraphics[width=\columnwidth]{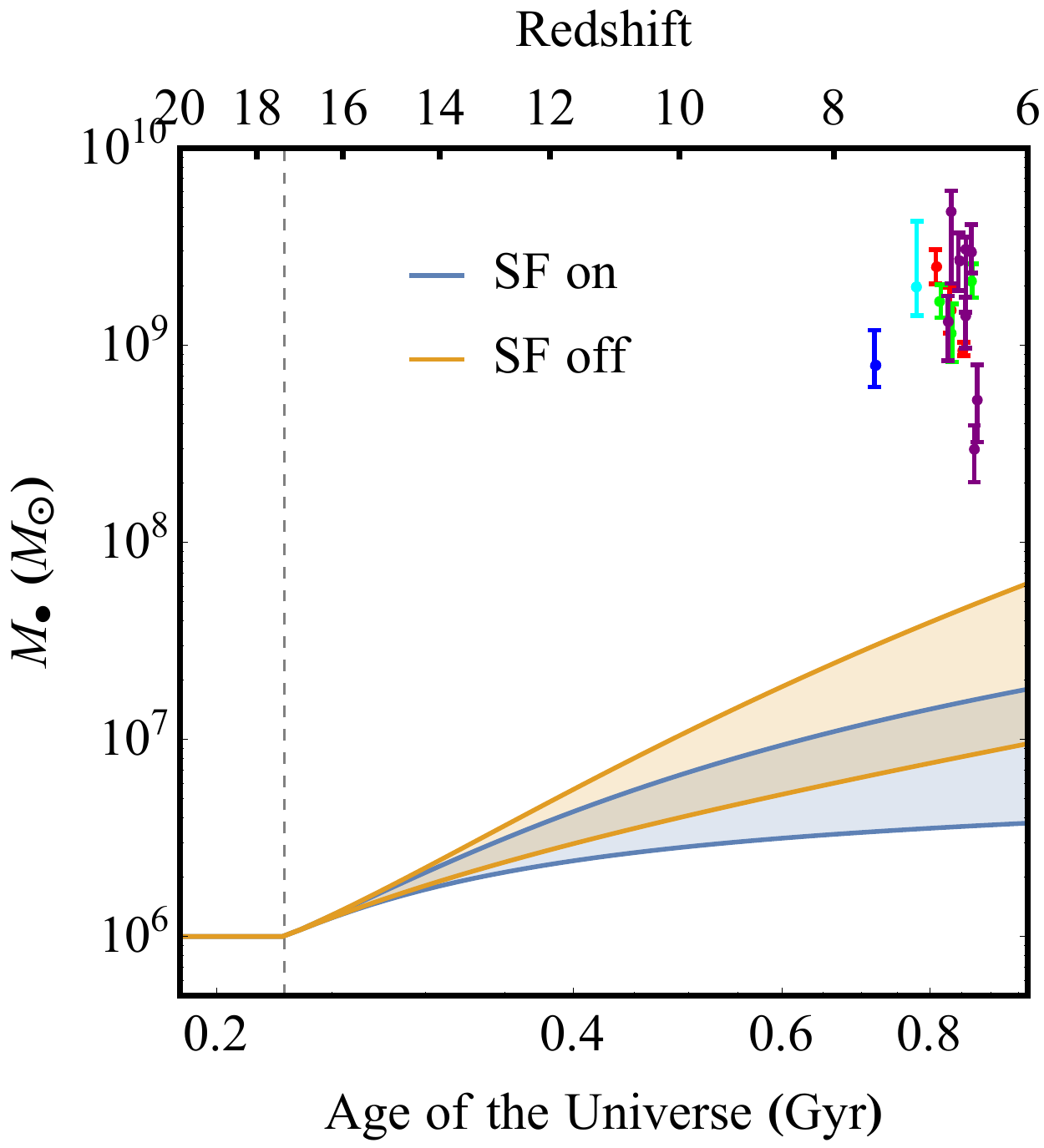}&
  \includegraphics[width=\columnwidth]{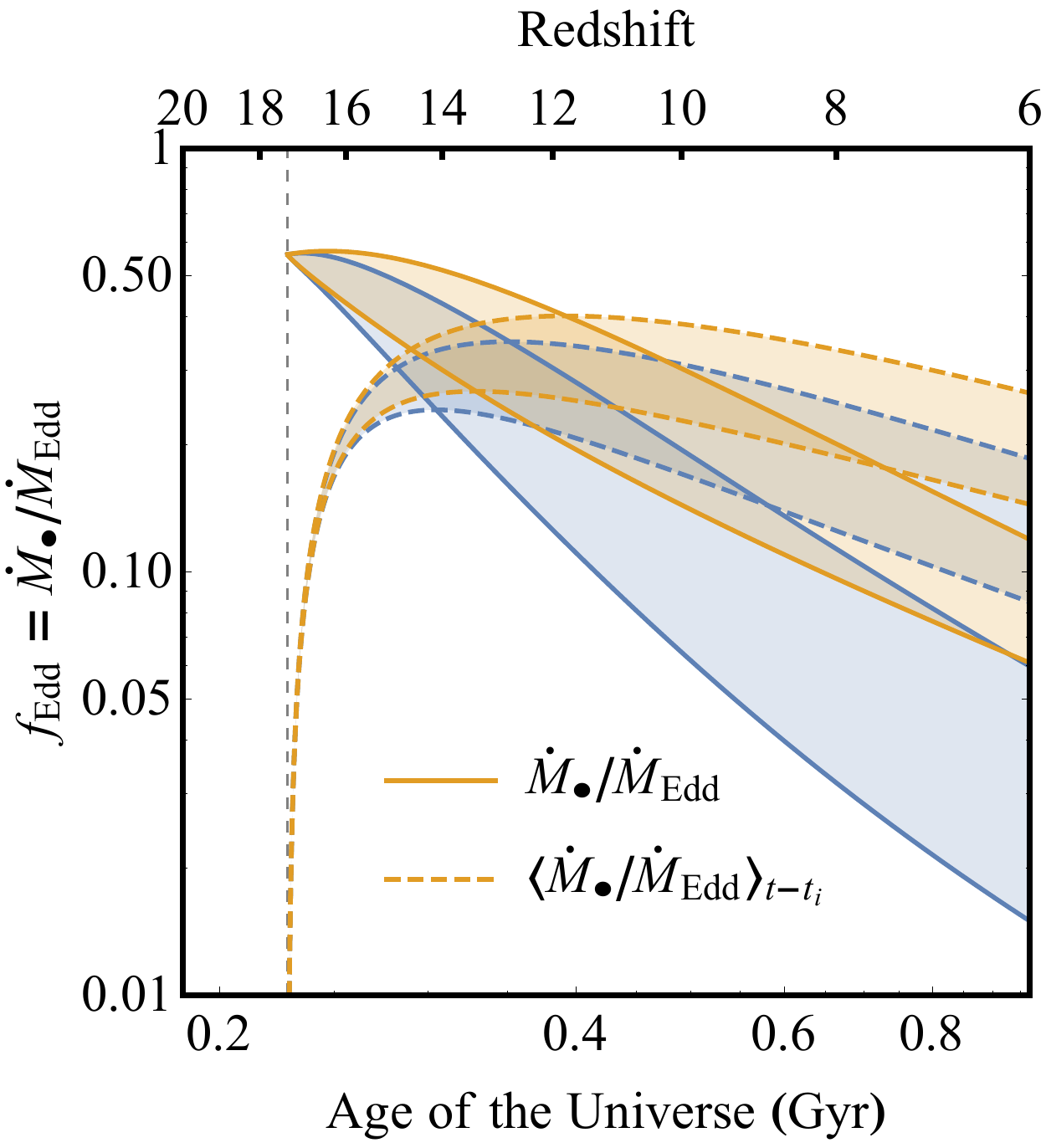}
\end{tabular}
    \caption{As in Figure~\ref{fig:mevo_mseed} except the evolution for a $M_{\rm seed} = 10^6 \, \mathrm{M_\odot}$ seed is shown for both with and without star formation included in the model (corresponding to SF on and SF off, respectively). The other model parameters are set to their fiducial values (see Table~\ref{tab:para_table}).}
    \label{fig:mevo_SF}
\end{figure*}

\subsection{Inflow rate in the disc}

The inflow rate is strongly dependent on both the absolute and relative masses of the disc and black hole. Figure~\ref{fig:mdotinf_mdmbh} shows the optimistic and conservative inflow rate estimates as a function of the disc and black hole masses. The dark matter halo mass scales with the total baryon mass ($M_{\rm b}=M_{\rm d} + M_{\bullet}$) so that the baryon fraction $f_{\rm b}=0.17$ holds. The optimistic inflow rate estimate (top left panel of Figure~\ref{fig:mdotinf_mdmbh}) strongly depends on the disc mass while the conservative is more dependent on the black hole mass (top right). On each panel, the dashed black line shows where $Q_{\rm min}=1$; below this line the disc is fully stabilised and no inflow can occur. At high black hole masses ($M_\bullet \gtrsim M_{\rm d}$), the disc becomes influenced by the gravitational potential of the black hole and is therefore more stable. This means the mass of the disc required for inflow to be possible is higher at high black hole masses. Indeed the optimistic inflow rate becomes more dependent on the black hole mass at higher black hole masses. However, as the inflow rate is defined at $R=R_{Q_{\rm min}}\sim R_{\rm d}$ and the black hole's sphere of influence is generally small ($R_\bullet << R_{\rm d}$), the disc mass largely determines the inflow rate. The conservative inflow rate estimate is more dependent on the black hole mass as the radius $R_{ \rm c, in}$ is closer to the black hole and thus more influenced by the central mass.

The red lines in the top panels of Figure~\ref{fig:mdotinf_mdmbh} show where the disc inflow rate is equal to 0.1, 1, and 10 times the Eddington accretion limit of the black hole. To the left of the $\dot{M}_{\rm inf}/\dot{M}_{\rm Edd} = 1 $ line the inflow rate is greater than the Eddington limit. To the right, the inflow rate is lower than the Eddington limit and thus supply to the reservoir the limiting factor in the growth of the black hole. Indeed, if a system is consistently to the right of the Eddington limit line (as will be the case for massive seeds with $M_{\rm seed} \sim 10^6  \, \mathrm{M_\odot}$), the Eddington limit will not be reached during the evolution. For lower mass seeds ($M_{\rm seed} \sim 10^3  \, \mathrm{M_\odot}$), the system will initially be to the left of this line but as the black hole will become more massive at later times, its growth will also be eventually limited by the growth of the reservoir. This indicates that for higher black hole accretion rates ($\dot{M}_\bullet \gtrsim \dot{M}_{\rm Edd}$) the inflow rate onto the gas reservoir is the more significant factor in determining the growth of a seed black hole, rather than the black hole accretion timescale.

\subsection{Black hole seed masses}

Figure~\ref{fig:mevo_mseed} shows the evolution of the fiducial model but with two different black hole seed masses $M_{\rm seed}=10^6 \, \mathrm{M_\odot}$ and $M_{\rm seed}=100 \, \mathrm{M_\odot}$. In each case the growth of the black hole is not capped at the Eddington rate and the growth rate of the black hole is effectively independent of the initial seed mass. Indeed, the black hole masses are roughly comparable following an initial period of super-Eddington accretion in the case with the lower mass seed. It is important to note in this case that, had the Eddington limit been implemented, it would not have changed the final black hole mass at $z=6$. The time-averaged Eddington fraction is calculated using the following:

\begin{equation}
\langle \dot{M}_\bullet /\dot{M}_{\rm Edd}\rangle_{t - t_{\rm i}} = \ln{\left [\frac{M_\bullet}{M_{\rm seed}} \right ]} \frac{t_{\rm Sal}}{t-t_{\rm i}}
\label{eddave}
\end{equation}

In both cases, $\langle \dot{M}_\bullet /\dot{M}_{\rm Edd}\rangle_{t - t_{\rm i}} < 1$  at the end of the calculation (shown by the dashed-line bounded regions in Figure~\ref{fig:mevo_mseed}c). As we do not model feedback effects, this means the final black hole masses at $z=6$ would have been the same had the black hole growth been capped at the Eddington limit.

The black hole seed mass does not play a role in determining the long term growth history of the black hole in our model. However, this ignores the effects of AGN feedback. With feedback effects included, the growth of the black hole would be more stunted due to the resulting outflows \citep{Johnson2011, Latif2018}. This could be particularly dramatic in the lower mass seed case as relatively more rapid growth is required and this would drive more energetic feedback.

\begin{figure*}
\centering
\begin{tabular}{lr}
  \includegraphics[width=\columnwidth]{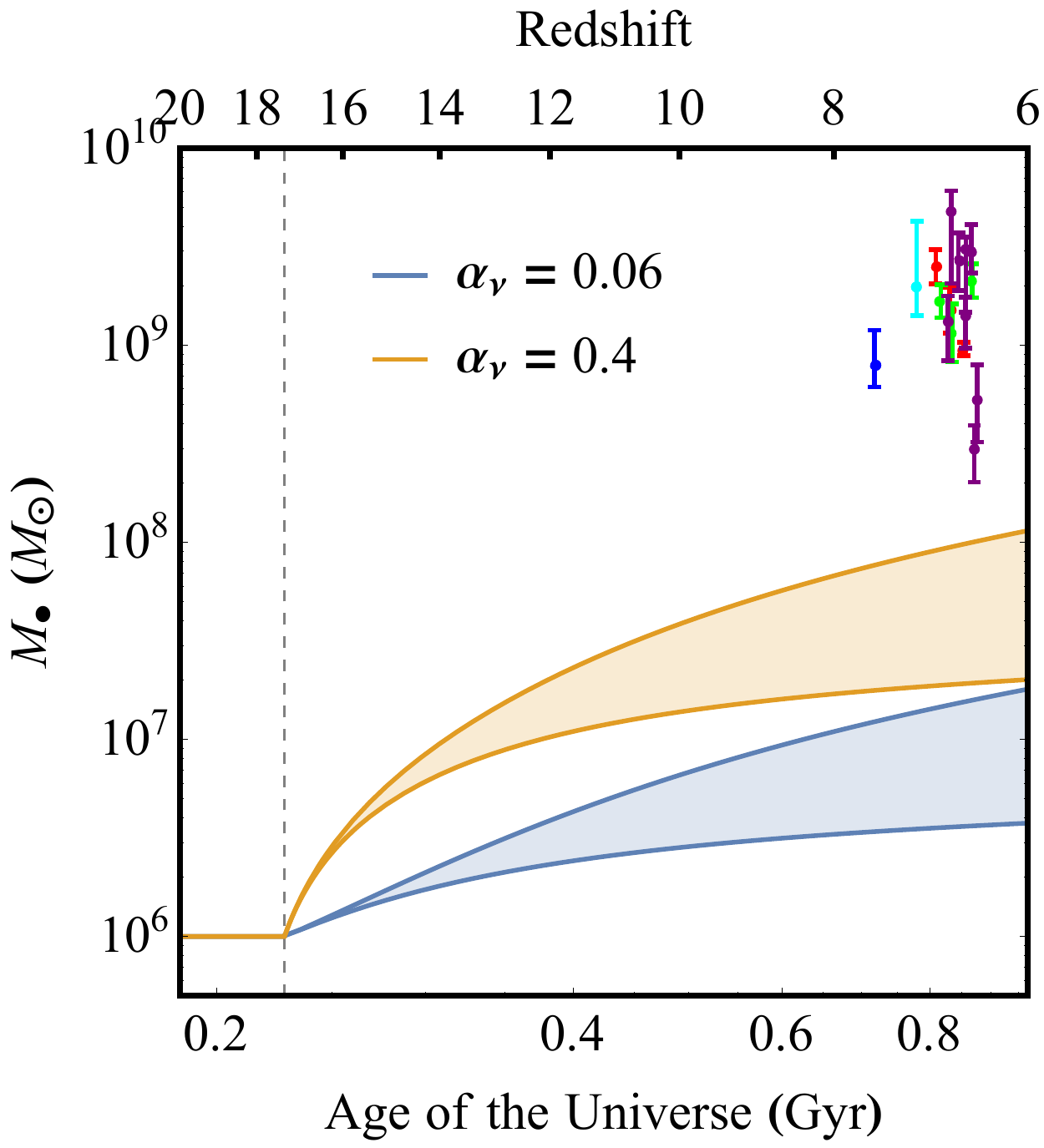}&
  \includegraphics[width=\columnwidth]{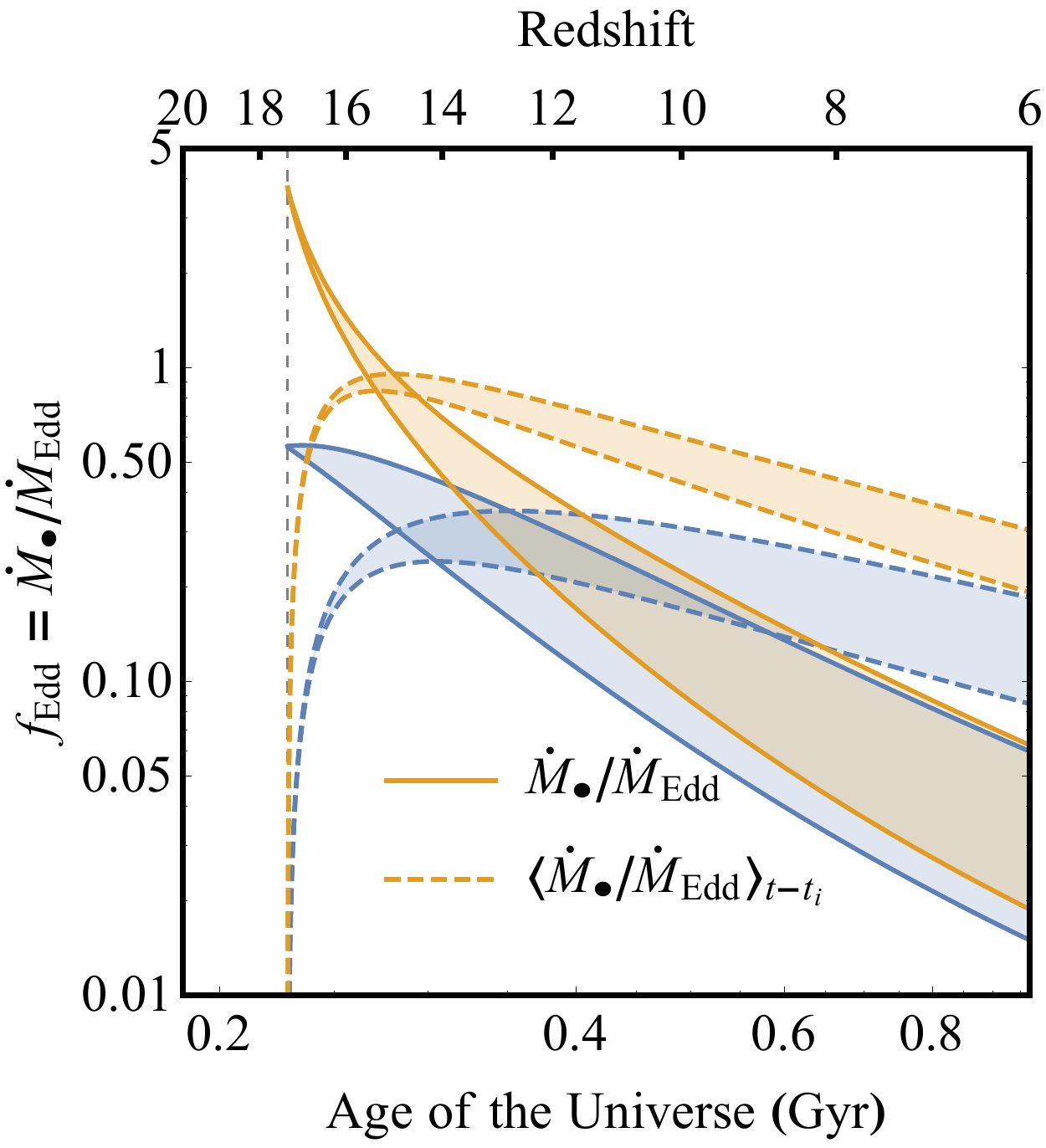}
\end{tabular}
    \caption{As in Figure~\ref{fig:mevo_mseed} except the evolution for a $M_{\rm seed} = 10^6 \, \mathrm{M_\odot}$ seed is shown with two values for the disc viscosity parameter, $\alpha_\nu=0.06$ and $\alpha_\nu=0.4$, which determines the inflow rate onto the gas reservoir. The other model parameters are set to their fiducial values (see Table~\ref{tab:para_table}).}
    \label{fig:mevo_alphavar}
\end{figure*}

\subsection{Star formation}

Star formation has the effect of slowing down the growth of the black hole. Though we do not model any feedback effects, star formation and black hole growth are both fed from the gas content of the disc and they are therefore in competition over the same gas supply. Figure~\ref{fig:mevo_SF} shows how the black hole growth in our fiducial case is changed when star formation is not included. When star formation is ``turned off", the final mass of the black hole increases as the higher gas fraction means the total mass that reaches the black hole is higher. After the initial onset of star formation in the fiducial case, the case with no star formation maintains a higher black hole growth rate. This difference in the growth rate is first noticeable around $z=15$ and as the model evolves it increases as the stellar disc fraction in the fiducial case increases to its highest value at $z=6$ of around $M_\star/M_{\rm d}\sim 0.9$. Even in the absence of star formation the final black hole mass is only $M_\bullet = 6.18 \times 10^7 \, \mathrm{M_\odot}$. The stellar mass (and stellar disc fraction) at $z=6$ varies with the growth rate of the halo \citep[see,][for more of a discussion]{Eastwood2018a}. If we take the extreme case where star formation is turned off in the model and chose our parameters to create a best case scenario for black hole growth (taking the maximum inflow rate with $\alpha_\nu=0.4$ with the earliest DCBH formation redshift $z_{\rm i}=20$ and the corresponding maximum halo growth rate $\zeta=0.926$), the black hole reaches a final mass of $M_\bullet = 6.57 \times 10^8\,\mathrm{M_\odot}$ with $\langle \dot{M}_\bullet /\dot{M}_{\rm Edd}\rangle_{t - t_{\rm i}} = 0.42$.

\subsection{Viscosity limited black hole growth}

With the optimistic inflow rate in our fiducial case, the final black hole mass at $z=6$ is $M_\bullet = 1.80 \times 10^7 \, \mathrm{M_\odot}$, more than 50 times smaller than the $M_\bullet\approx 10^9\, \mathrm{M_\odot}$ target.

Increasing the viscosity parameter of the disc does increase the inflow rate and thus black hole growth rate. Figure~\ref{fig:mevo_alphavar} compares the evolution of the black hole masses for two values of the viscosity parameter, $\alpha_\nu=0.06$ and $\alpha_\nu=0.4$. With the higher disc viscosity the black hole is fed more rapidly and therefore can reach a higher mass estimate of $M_\bullet = 1.14 \times 10^8 \, \mathrm{M_\odot}$. However, as discussed above, it is not clear that this higher viscosity parameter is applicable for galaxy discs as clumps are expected to form for $\alpha_\nu>0.06$. Again in this scenario the Eddington limit does not affect the final mass of the black hole as $\langle \dot{M}_\bullet /\dot{M}_{\rm Edd}\rangle_{t - t_{\rm i}} < 1$ at the end of the calculation (shown by the dashed-line bounded regions in the right panel of Figure~\ref{fig:mevo_alphavar}) for either value of $\alpha_\nu$.

\begin{figure*}
\centering
  \begin{minipage}[t]{0.32\textwidth}
    \vspace{0pt}\includegraphics[width=\hsize, keepaspectratio=true]{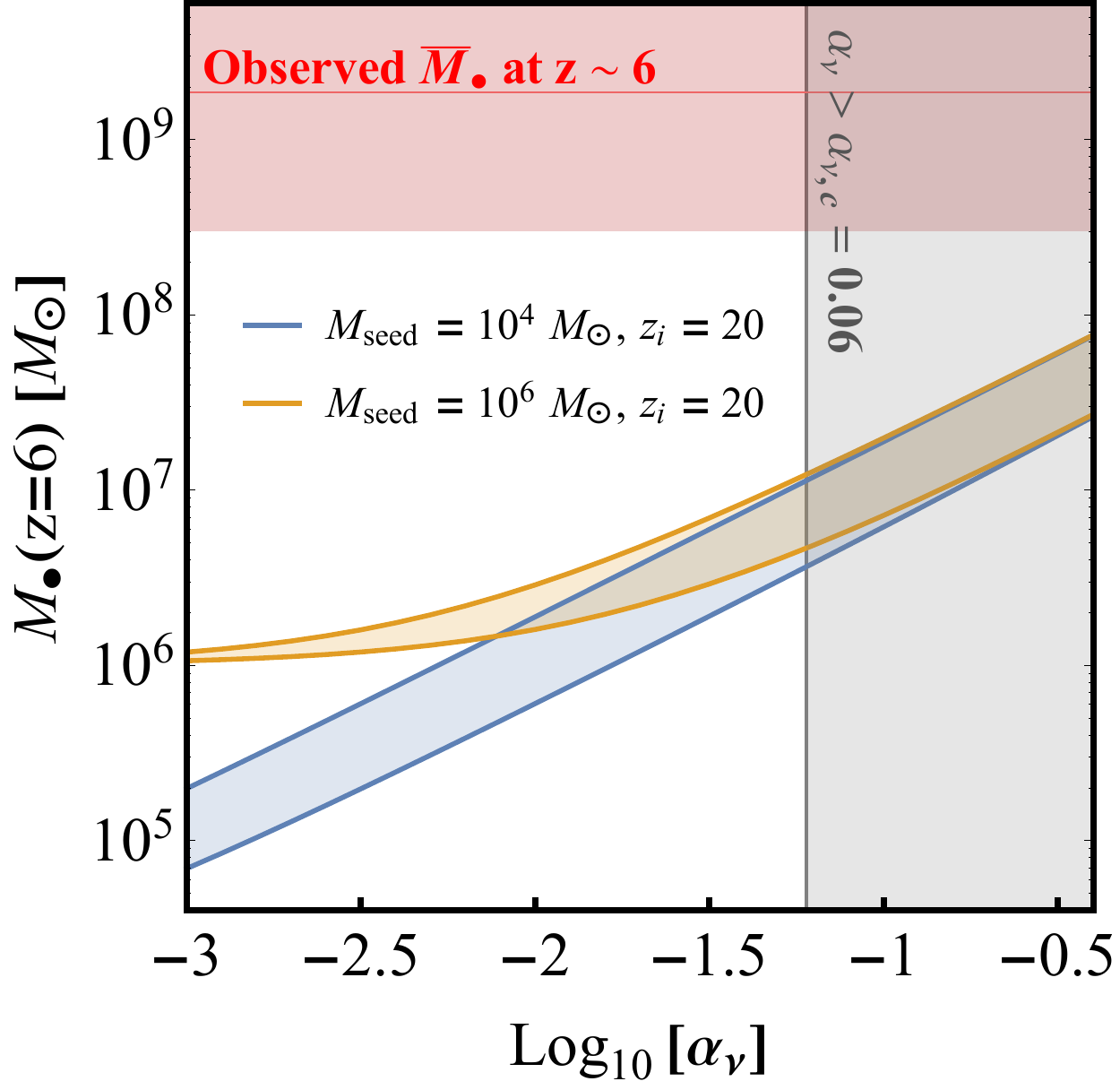}
  \end{minipage}
  \hfill
  \begin{minipage}[t]{0.32\textwidth}
    \vspace{0pt}\includegraphics[width=\hsize, keepaspectratio=true]{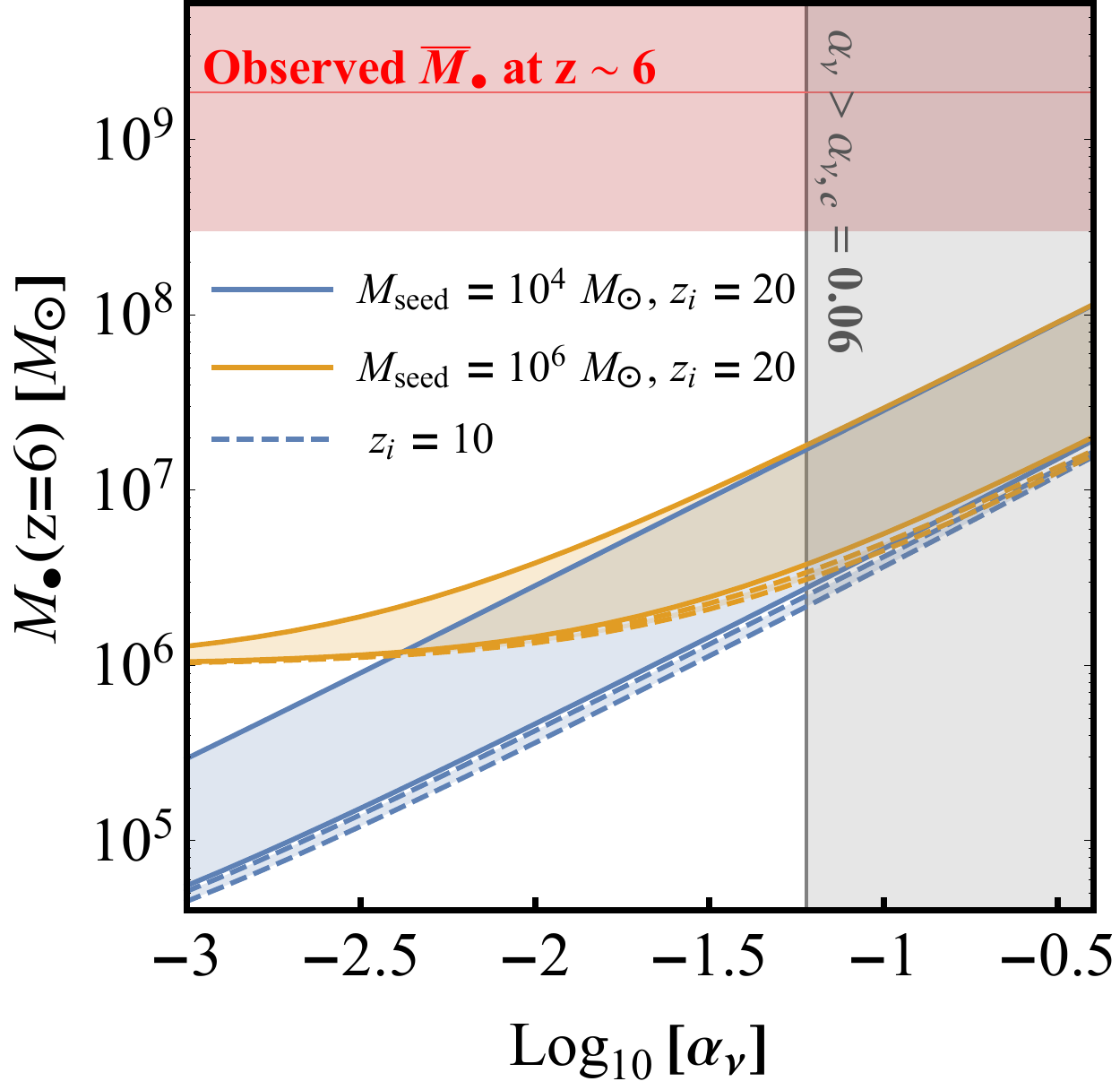}
  \end{minipage}
  \hfill
  \begin{minipage}[t]{0.32\textwidth}
    \vspace{0pt}\includegraphics[width=\hsize, keepaspectratio=true]{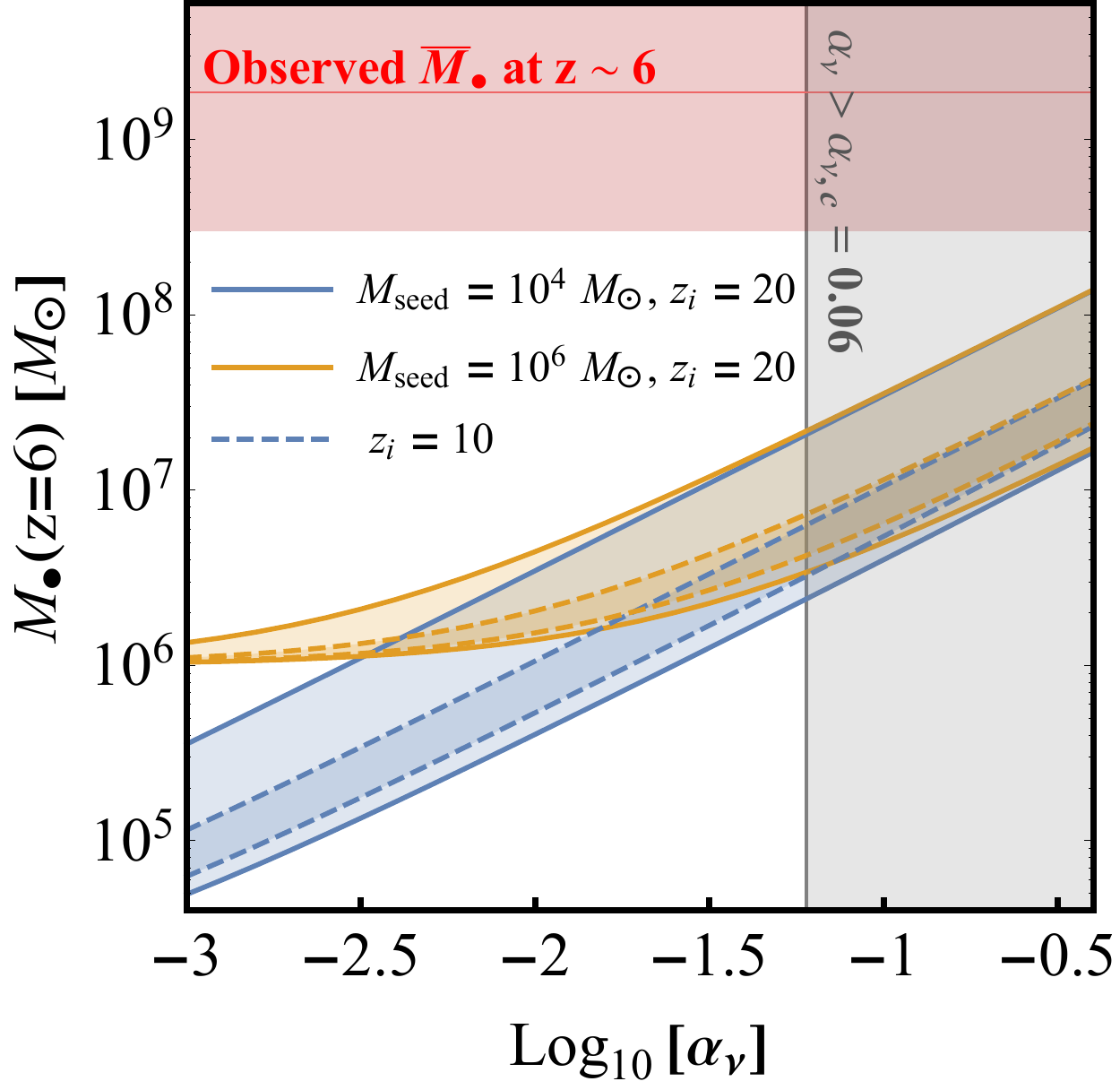}
  \end{minipage}
    \begin{minipage}[t]{0.32\textwidth}
    \begin{center}
    \textbf{(a)}
    \end{center}
  \end{minipage}
  \hfill
  \begin{minipage}[t]{0.32\textwidth}
    \begin{center}
    \textbf{(b)}
    \end{center}
  \end{minipage}
  \hfill
  \begin{minipage}[t]{0.32\textwidth}
    \begin{center}
    \textbf{(c)}
    \end{center}
  \end{minipage}
    \caption{The mass of the black hole at $z=6$ as a function of the disc viscosity parameter, $\alpha_\nu$. Each panel shows the case for a different halo growth rate parameter, $\zeta$ (a) $\zeta=0.568$, (b) $\zeta=0.806$, and (c) $\zeta=\zeta_{\rm max}$ ($\zeta_{\rm max}=0.926$ for $z_{\rm i}=20$ and $\zeta_{\rm max}=1.462$ for $z_{\rm i}=10$). The possible final black hole masses are shown as the shaded regions between the curves calculated using the upper and lower estimates for the inflow rate onto the gas reservoir. The solid and dashed lines indicate the cases where the seed formation redshift is $z_{\rm i}=20$ and $z_{\rm i}=10$ respectively. The greyed out region at $\alpha_\nu>0.06$ indicates the region where disc instabilities will limit the inflow rates significantly \citep{Rice2005}. The red region indicates the range of black hole mass observed at $z>6$ with the solid pink line indicating the average of these masses (based on data from \citet{Mortlock2011}, \citet{DeRosa2014}, \citet{Mazzucchelli2017}, \citet{Banados2018} and \citet{Reed2019arXiv}). The remaining model parameters are set to their fiducial values (see Table~\ref{tab:para_table}).}
    \label{fig:mbhmax_alpha}
\end{figure*}

Figure~\ref{fig:mbhmax_alpha} shows how the final mass of the black hole varies with $\alpha_\nu$ for three different halo growth rates ($\zeta=0.568$, $\zeta=0.806$, and $\zeta=\zeta_{\rm max}$), two seed masses ($M_{\rm seed}= 10^4\, \mathrm{M_\odot}$ and $M_{\rm seed}= 10^6\, \mathrm{M_\odot}$), and two seed formation redshifts ($z_{\rm i}=20$ and $z_{\rm i}=10$). $\zeta_{\rm max}$ varies with the seed formation redshift so that the $\zeta_{\rm max} = 0.926$ for $z_{\rm i}=20$ and $\zeta_{\rm max} = 1.462$ (1-$\sigma$ above the average halo growth rate at $z=6$) for $z_{\rm i}=10$. At low values of the viscosity parameter ($\alpha_\nu < 0.005$), there is a significant difference in the final black hole mass for the different initial seed masses. This is simply because the growth of the black holes is weak, in the case of the highest initial seed mass, the seed takes up a significant fraction of that final black hole mass. This difference disappears in the higher halo growth rate case as the inflow rates are much higher. At higher $\alpha_\nu$ the final mass is independent of the seed mass as the higher inflow rate means that the total mass accreted onto the reservoir (and then onto the black hole) is much greater than the initial mass of the black hole for either seed.

In the case with the lowest halo growth rate no $z_{\rm i}=10$ curves are shown as the disc is never unstable in this scenario and therefore no inflow occurs. At higher halo growth rates ($\zeta=0.806$, $\zeta=0.926$ and $\zeta=1.462$) we can see that the seed formation redshift is important in determining the maximum mass to which the black hole can grow. The final mass of the black hole is much lower for the $z_{\rm i}=10$ cases. This is not only because there is much less time for the black hole to grow but, significantly, the halo mass at $z=6$ decreases for lower $z_{\rm i}$ and thus the growth rate of the disc decreases. This comes from the halo mass at the seed formation redshift being set as the mass of an atomic-cooling halo with $T_{\rm vir}=8000$ K at that epoch before growing at the rate determined by Equation~\ref{halogrowth}.

The red region on each panel in Figure~\ref{fig:mbhmax_alpha} indicates the range of masses of the quasars observed at $z>6$ and the pink line indicates the average of these masses \citep{Mortlock2011,DeRosa2014,Mazzucchelli2017,Banados2018,Reed2019arXiv}. In all cases the model cannot reach the observed mass range, even with $\alpha_\nu = 0.4$.

\subsection{Cosmic accretion limited growth}

\begin{figure*}
\centering
\begin{tabular}{lr}
  \includegraphics[width=\columnwidth]{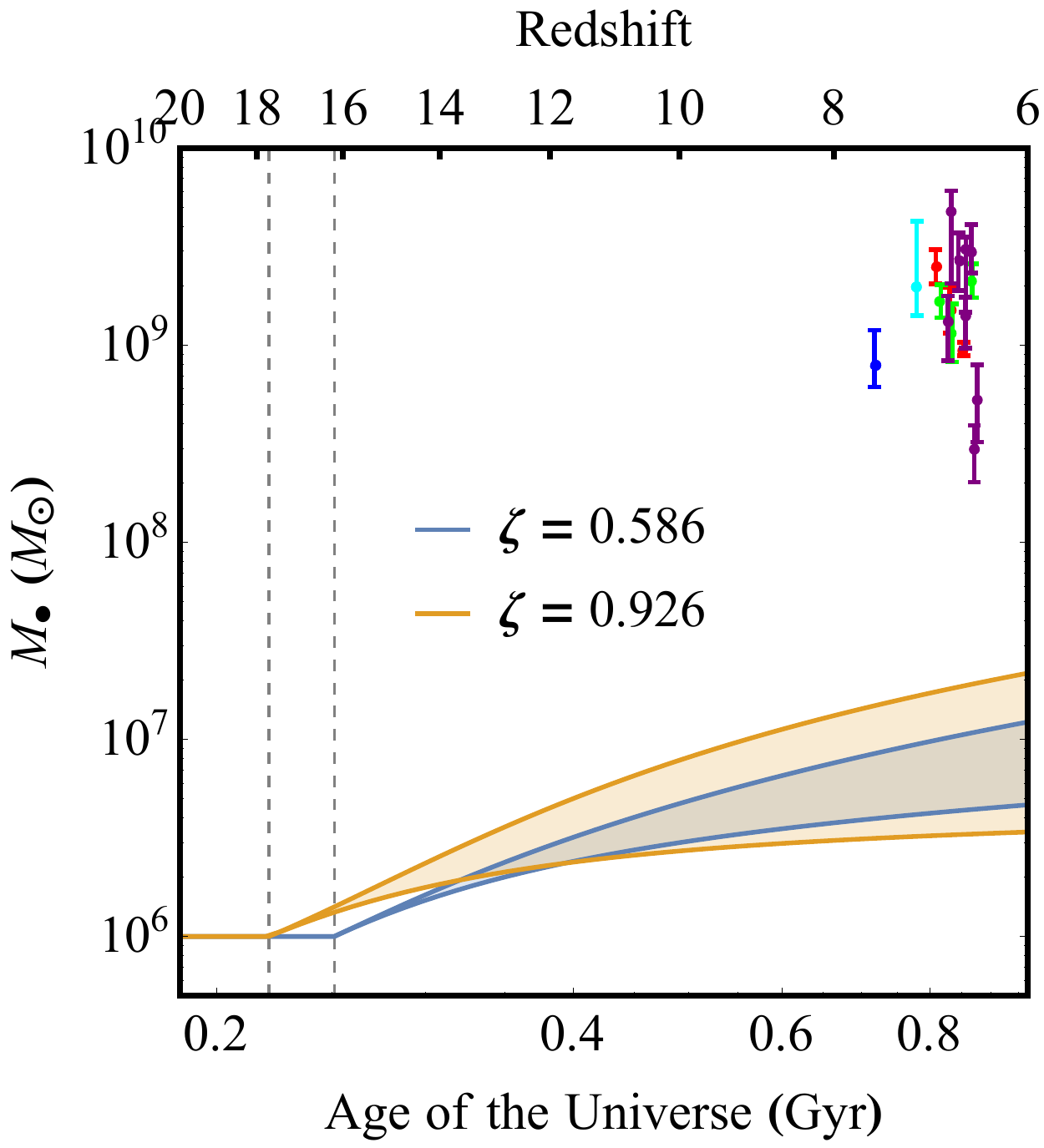}&
  \includegraphics[width=\columnwidth]{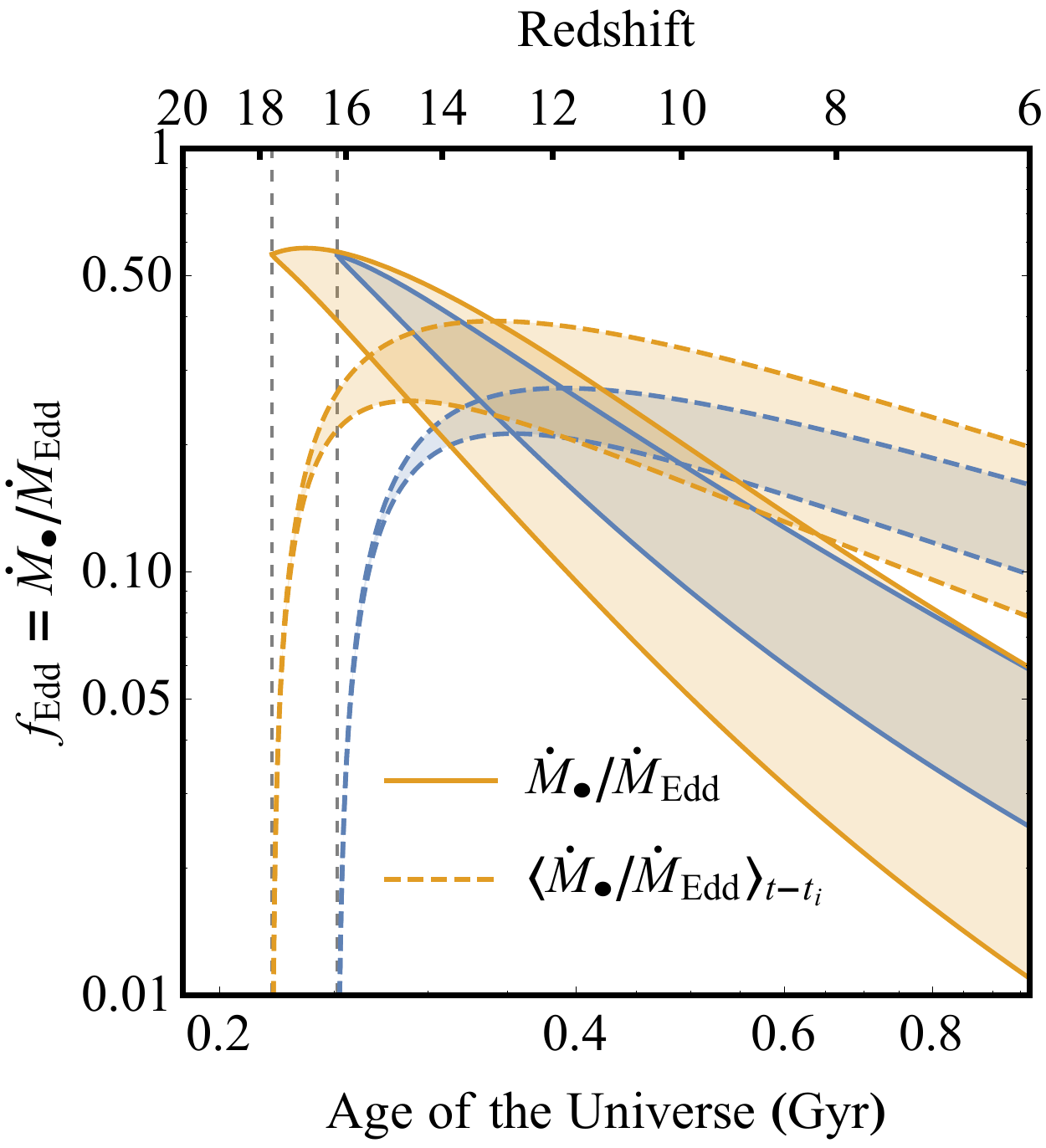}
\end{tabular}
    \caption{As in Figure~\ref{fig:mevo_mseed} except the evolution for a $M_{\rm seed} = 10^6 \, \mathrm{M_\odot}$ is shown for two halo growth rate parameters, $\zeta=0.586$ and $\zeta=0.926$. The lower value of $\zeta$ mimics the growth rate of modelled DCBH hosting subhaloes \citep{Agarwal2016a}. The upper value corresponds to the growth rate required to grow atomic cooling haloes at the formation redshift $z_{\rm i} = 20$ into $M\sim 10^{13} \, \mathrm{M_\odot}$ haloes at $z=6$. Haloes with this mass or above at $z=6$ have an abundance of $\phi \sim1\,\mathrm{Gpc^{-3}}$ \citep{Sheth2001, Murray2013}, matching the SMBH abundance \citep{Fan2003}. The other model parameters are set to their fiducial values (see Table~\ref{tab:para_table}).}
    \label{fig:mevo_zetavar}
\end{figure*}

The growth rate of the halo determines the rate at which the disc can grow. Increasing the halo growth rate parameter will increase the accretion rate onto the disc. This in turn will increase the disc mass, making the disc more unstable and increase the resulting inflow rate in the disc. For a given formation redshift, $z_{\rm i}$, increasing the halo growth rate will increase the final halo mass at $z=6$. Figure~\ref{fig:mevo_zetavar} shows evolution of the black hole and gas reservoir for two different values of the halo growth rate parameter, $\zeta=0.586$ and $\zeta=0.926$. In the $\zeta=0.586$ case the lower halo growth results in a more stunted growth of the black hole with the maximum black hole mass estimate (using the optimistic inflow rate) at $z=6$ is $M_\bullet = 1.22 \times 10^7 \mathrm{M_\odot}$. With a lower halo growth rate the star formation is less significant and therefore the conservative black hole estimate is not as significantly stunted. At $\zeta=0.926$, the disc mass grows more rapidly resulting in a higher inflow rate onto the reservoir and a higher growth rate of the black hole. The resulting maximum black hole mass estimate at $z=6$ is almost a factor of 2 larger at $M_\bullet = 2.17 \times 10^7 \mathrm{M_\odot}$.

The difference in minimum black hole mass estimates (corresponding to the conservative inflow rate) in Figure~\ref{fig:mevo_zetavar} is a result of the different stellar disc fractions. Increasing the growth rate of the halo will increase the disc mass at a given time, however, as the inflow rate does not change dramatically with the disc mass, the inflow rate does not change significantly other than due to the difference in the stellar disc fraction. As this fraction is lower in the lower growth rate case the minimum black hole growth rate is higher.

\subsection{Halo growth rate versus viscosity parameter}

\begin{figure*}
\centering

\begin{tabular}{lr}
  \includegraphics[width=\columnwidth]{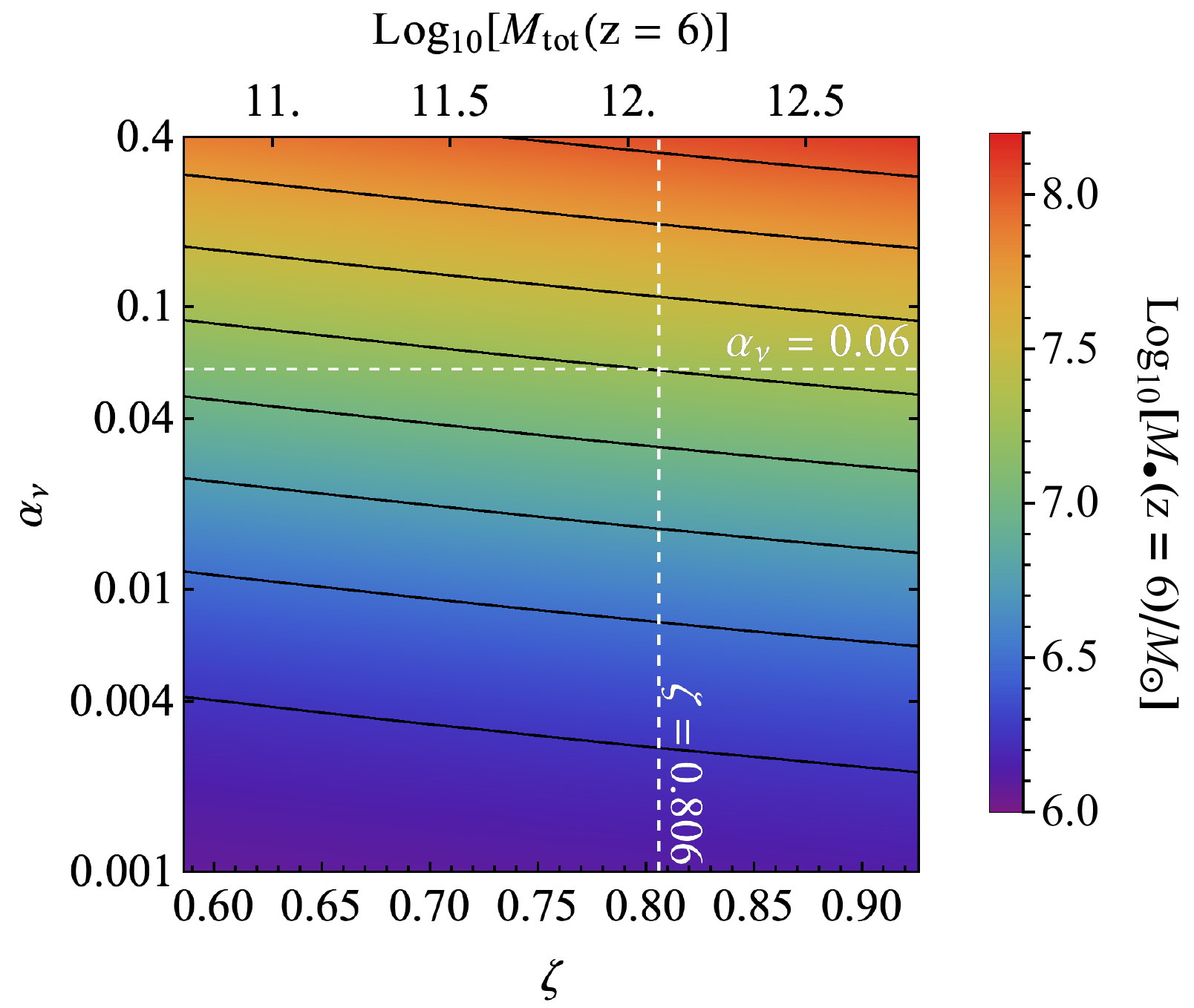}&
  \includegraphics[width=\columnwidth]{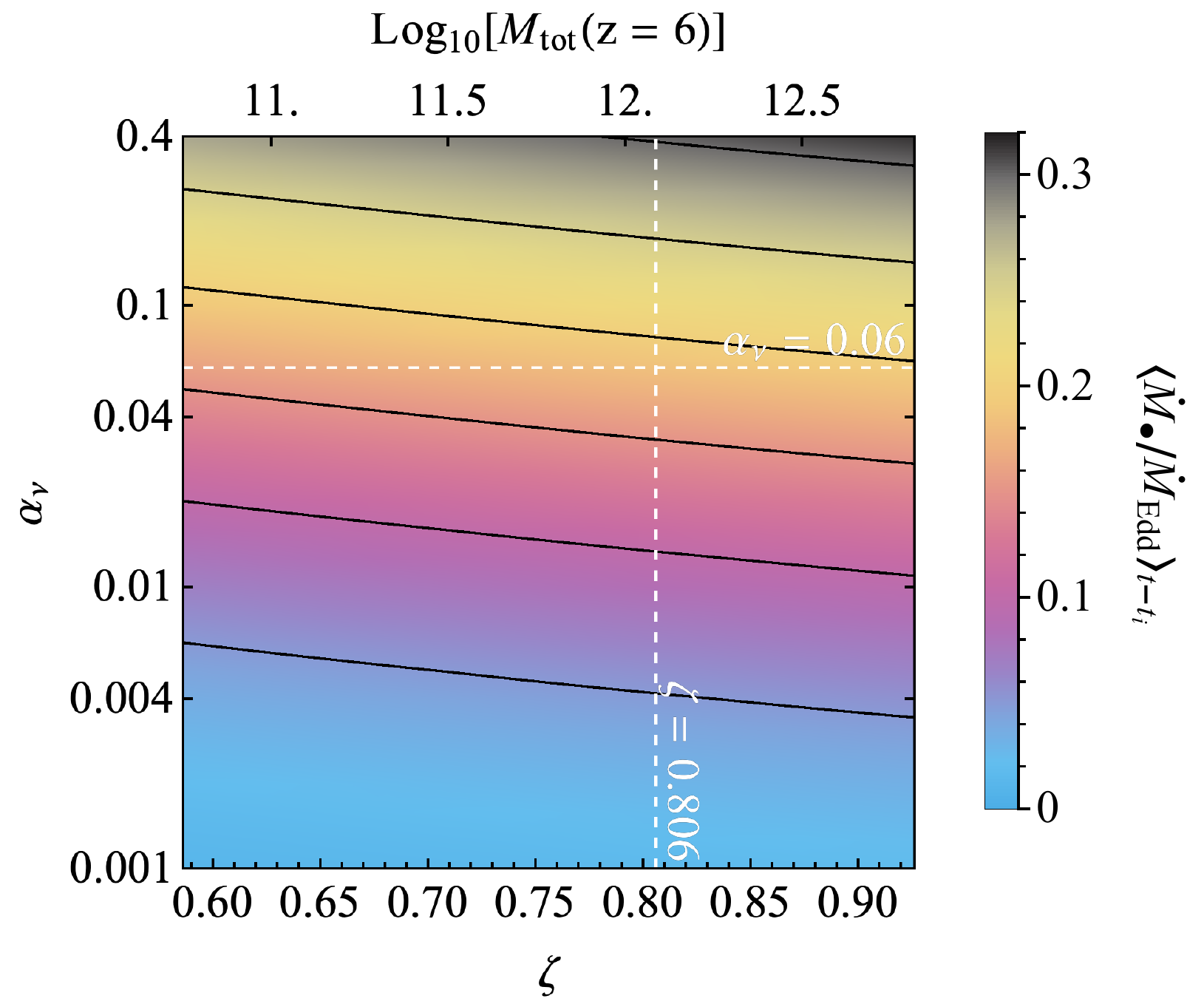}
\end{tabular}
    \caption{The left panel shows the mass of the black hole at $z=6$ as a function of the halo growth rate parameter, $\zeta$, and the viscosity parameter, $\alpha_\nu$. The initial seed, with mass $M_{\rm seed} = 10^6 \, \mathrm{M_\odot}$ forms at $z=20$ and is fed using the optimistic inflow rate. The right panel shows the mean Eddington fraction required for the black hole to grow to the final mass indicated by the left panel. The horizontal and vertical, white, dashed lines correspond to the critical viscosity value $\alpha_{\nu}=0.06$ \citep{Rice2005} and the average halo growth rate parameter value $\zeta=0.806$ \citep{Neistein2008}.}
    \label{fig:mbhmax_alpha_zeta}
\end{figure*}

The left panel of Figure~\ref{fig:mbhmax_alpha_zeta} shows for the case of the optimistic estimate of inflow rate how the final mass of the black hole at $z=6$ varies with $\alpha_\nu$ and $\zeta$ for a $M_{\rm seed}=10^6 \, \mathrm{M_\odot}$ formed at $z_{\rm i}=20$. As discussed previously, the final black hole mass increases with both $\alpha_\nu$ and $\zeta$. This results in a maximum value for $\zeta<0.926$ of $M_\bullet = 1.39 \times 10^8 \, \mathrm{M_\odot}$ ($M_\bullet = 6.57 \times 10^8 \, \mathrm{M_\odot}$ without star formation).

The right hand panel of Figure~\ref{fig:mbhmax_alpha_zeta} shows the time-averaged Eddington fraction for the black hole to grow from its seed mass to the final mass shown in the left hand panel. For $\zeta<0.926$, this fraction is has a maximum of $f_{\rm Edd}\sim0.32$ ($f_{\rm Edd}\sim0.42$ without star formation), indicating that the final mass is limited by the total mass accreted onto the reservoir rather than the Eddington limit or efficiency of the black hole growth.

Within the range of the parameters for halo growth and inflow rates investigated here, there are no cases (with star formation implemented) where the black hole mass at $z=6$ reaches the observed range of masses at this redshift with an average of $\bar{M}_\bullet = 1.85 \times 10^9 \, \mathrm{M_\odot}$ \citep{Mortlock2011, DeRosa2014, Mazzucchelli2017, Banados2018, Reed2019arXiv}. Note that if we neglect to include star formation and chose model parameters which create a best case scenario for black hole growth (taking the maximum, super-critical inflow rate with $\alpha_\nu=0.4$, the earliest DCBH formation redshift $z_{\rm i}=20$, and the corresponding maximum halo growth rate $\zeta=0.926$), the black hole reaches a final mass of $M_\bullet = 6.57 \times 10^8\,\mathrm{M_\odot}$. If we ignore the upper limit placed on the halo mass at $z=6$ by the observed SMBH abundance and use a halo growth rate parameter of $\zeta=1.462$, the black hole reaches a maximum mass of $M_\bullet = 2.63 \times 10^{8} \, \mathrm{M_\odot}$. Note this extreme case is only mentioned here to illustrate the limitations on black hole growth in the model and should not be considered physically viable as the halo becomes overly massive. That is, it reaches a mass of $\sim 10^{16} \, \mathrm{M_\odot}$, corresponding to a halo number density at $z=6$ many orders of magnitude below the observed SMBH abundance. From the halo mass function \citep{Sheth2001, Murray2013}, we can see that no halo of this mass should exist at this redshift in the observable universe. Indeed, only $\mathcal{O}(1)$ halo with a mass $\sim 10^{13.2} \, \mathrm{M_\odot}$ should exist in the observable universe at $z=6$.

\section{Merger driven black hole growth} \label{sect:mergers}


DCBHs are thought to form in close neighbours of massive star forming galaxies \citep[see, e.g.][]{Agarwal2014a, Wise2019} that subsequently become satellites. This proximity means the DCBH host will eventually undergo a merger with a neighbouring halo, having a major impact on the evolution of the black hole \citep{Dijkstra2008}.

Galaxy merger events can be a major driver of angular momentum transport and therefore accretion \citep{DOnghia2006}. Such events could therefore be responsible for driving gas accretion. Here we aim to estimate the maximum capability of mergers to feed black holes.

\citet{Hopkins2009} argue that merger driven bar instabilities could cause the efficient loss of angular momentum for gas within the radius

\begin{equation}
    \frac{R_{\rm mer}}{R_{\rm d}} \lesssim \left (1 - f_{\rm gas} \right) f_{\rm disc} \, \frac{2\mu}{1+\mu} \, F(\theta, b,\mu)
    \label{eq:rmerge}
\end{equation}

\begin{equation}
    F(\theta, b, \mu) \equiv \frac{\left(1+\left(b/R_{\rm d} \right)^2 \right)^{-3/4} }{\left(1+\left(b/R_{\rm d} \right)^2 \right)^{3/4} -\sqrt{2(1+\mu)} \cos{\theta}} 
    \label{eq:factor-rmerge}
\end{equation}

\noindent where $\mu$ is the merger mass ratio, $b$ is the pericentric distance on the relevant final passage, $\theta$ is the inclination of the orbit relative to the disc plane, $f_{\rm disc}$ is the disc mass fraction of the total mass enclosed in $R_{\rm d}$ and $f_{\rm gas}$ is the gas fraction in the disc. The dependence of $R_{\rm mer}$ on the merger orbital parameters, $F(\theta, b, \mu)$, varies strongly. As $\theta$ and $b$ are not determined by our model, we take the mean value of $F(\theta, b, \mu) \sim 1.2$ from \citet{Hopkins2006} to calculate a typical disruption effect of a merger and as a maximum we assume the entire gas disc is disrupted (this scenario can be reached for different values of $\theta$, $b$, and $\mu$).

The gas enclosed in $R_{\rm mer}$ will become unstable and collapse efficiently towards the centre of the system with some fraction going into growing the reservoir which feeds the black hole. Here we take the extreme case where this fraction is unity and there is no starburst following the merger to give an upper limit on the possible black hole growth.

When a merger event occurs, $R_{\rm mer}$ is calculated using equations~\ref{eq:rmerge} and \ref{eq:factor-rmerge} and the gas mass inside this radius is taken from the disc and added to the gas reservoir over a dynamical time (taken as $t_{\rm dyn} =  1/\Omega$ at $R_{\rm d}$). During this time no star formation may occur.

\begin{figure}
\centering
  \includegraphics[width=\columnwidth]{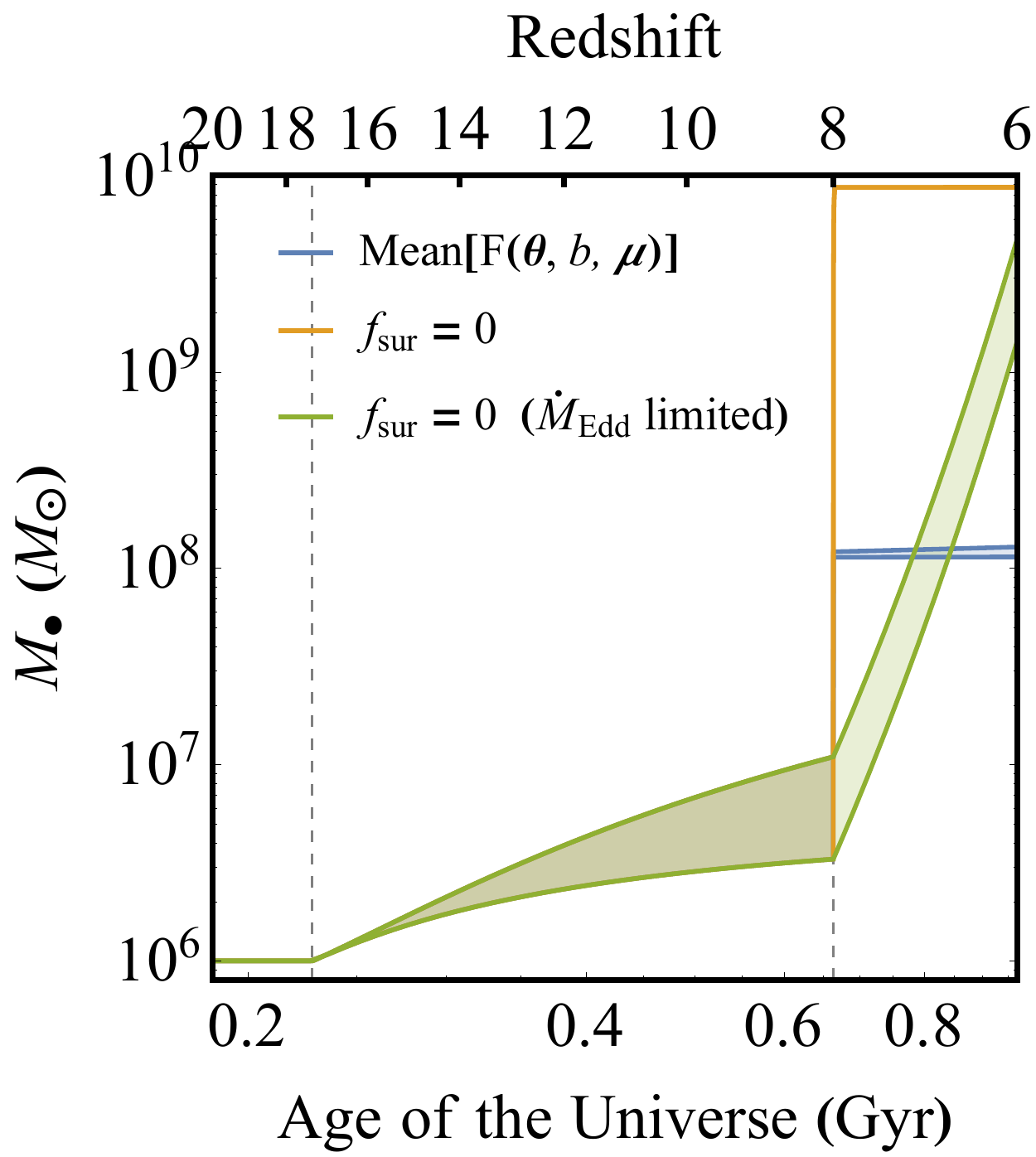}
    \caption{The mass evolution of the black hole for the two merger events modelled here. The orange region shows the case where the disc is fully disrupted (i.e. the survival fraction of the disc is $f_{\rm sur}=0$). The green region is the same ($f_{\rm sur}=0$) except with the black hole growth limited at the Eddington rate. The blue region shows the case with the typical value for $F(\theta, b, \mu)$ (see Equations~\ref{eq:rmerge} and \ref{eq:factor-rmerge}). In each case the merger is set to occur at $z=8$. The black hole masses are shown as the shaded regions between the curves calculated using $\dot{M}_{\rm inf}(R_{Q_{\rm min}})$ and $\dot{M}_{\rm inf}(R_{\rm c, in})$ as the upper and lower estimates for the inflow rate onto the gas reservoir respectively. The vertical dashed lines represent, from left to right, the time when $Q_{\rm min}=1$ and the time of the merger event. The model parameters in Table~\ref{tab:para_table} are set to their fiducial values.}
    \label{fig:mevo_merger}
\end{figure}

For the parameter space investigated here, the stellar component of the disc reaches a maximum fraction of the disc mass of $f_{\star}\sim0.9$. In this case the radius inwards of which the disc is disrupted by a typical merger event is small $R_{\rm mer}/R_{\rm d}\lesssim 0.2$. As a fraction of the total disc mass, the mass disrupted is only $\sim 1 \%$ (giving a survival fraction of the disc of $f_{\rm sur}=0.99$). However, the disruption radius is strongly dependent on the merger orbital parameters, $\theta$ and $b$, and the entire disc can be disrupted ($f_{\rm sur}=0$) in one $1:1$ merger event. Figure~\ref{fig:mevo_merger} shows the evolution of the black hole mass for the two cases of mergers. For the typical merger, the disruption causes a jump in the available mass in the reservoir and the black hole mass rapidly increases by at least an order of magnitude to $M_\bullet \sim 10^8 \, \mathrm{M_\odot}$. In the $f_{\rm sur}=0$ case, however, the jump is much more dramatic with the black hole mass reaching $M_\bullet \sim 10^{10} \, \mathrm{M_\odot}$ if the growth is not limited by the Eddington rate. If the Eddington limit is introduced, the black hole mass at $z=6$ is estimated between $M_\bullet \sim 1.5 \times 10^{9} \, \mathrm{M_\odot}$ and $M_\bullet \sim 5 \times 10^{9} \, \mathrm{M_\odot}$.  This is an extreme case and is only shown to illustrate how a strong merger event can efficiently drive the growth of SMBHs.

\section{Discussion \& Summary} \label{sect:summary}

In this study we analytically model the idealised growth of massive seed black holes via the inflow of gas from growing proto-galactic discs. The inflow rate of gas in the disc is a function of the disc gravitational stability (Equation~\ref{mdotinf2}) and thus depends on the masses of both the disc and the black hole (Figure~\ref{fig:mdotinf_mdmbh}). We find that for a typical host halo, black growth is limited by the inflow rate, and that even in the absence of feedback effects high Eddington ratios $\sim 1$ will not be reached. Indeed, for our fiducial case we find an upper black hole mass estimate of $M_\bullet = 1.80 \times 10^7 \, \mathrm{M_{\odot}}$ (Figure~\ref{fig:mevo_mseed}), indicating that viscosity driven accretion is too inefficient to feed the growth of seeds into $M_\bullet \sim 10^9 \, \mathrm{M_\odot}$ SMBHs within the first billion years of the Universe.

If the growth rate of the black hole is not manually capped by the Eddington limit, we find that the initial seed mass of the black hole has limited to no bearing on the mass the black hole can reach by $z\sim6$ (Figure~\ref{fig:mevo_mseed}). However, in this calculation we have ignored the effects of feedback which drives outflows and can stunt black hole growth \citep{Johnson2011, Dubois2015, Latif2018}. With the inclusion of feedback effects and the resulting lower black hole accretion rate, the difference in the final mass of different seeds would likely be more significant.

We find that SMBHs can grow more easily in faster growing haloes, where the more massive discs that form are more gravitationally unstable and therefore inflow is stronger. Higher halo growth rates are expected in large $\sigma$-fluctuations of the cosmic density field in the Universe. However, the host halo masses in which a seed can grow to $M_\bullet\sim 10^9 \, \mathrm{M_\odot}$ are less abundant than SMBHs at $z\sim6$. When taking the $\phi \sim 1\,\mathrm{Gpc^{-3}}$ abundance of SMBHs \citep{Fan2003} into account in matching the corresponding host halo mass, we find a maximum black hole mass at $z=6$ of only $M_\bullet = 1.42 \times 10^8 \, \mathrm{M_{\odot}}$ (with $\alpha_\nu =0.4$ and $z_{\rm i}=20$, see Figure~\ref{fig:mbhmax_alpha}). This strongly implies that the observed population of SMBHs at $z=6$ did not grow steadily within isolated halos, and are not fed solely through viscosity driven inflow. Indeed, the physical process of black hole growth at high redshift is more complex. DCBHs are expected to form in haloes which will later become satellite subhaloes and will therefore likely experience low accretion rates. More rapid black hole growth should be expected during and following the merger of the DCBH host with a central galaxy. Major mergers provide the most promising avenue for SMBH growth through efficiently dissipating angular momentum and driving gas towards the black hole.

We modelled the inflow rate within the disc as viscosity driven accretion. The viscosity parameter, $\alpha_\nu$, was assumed to be a constant for each calculation, rather than calculating $\alpha_\nu$ as a function of the gravitational stability of the disc \citep[see, e.g.][]{Devecchi2010}. However, the fiducial value of $\alpha_\nu=0.06$ should be seen as an upper limit as higher values lead to disc fragmentation \citep{Rice2005}. Our estimates of the inflow rate due to viscosity (and as a result black hole growth rates and final black hole masses) can therefore be considered conservatively large.

The inflow rates calculated here could be further overestimated as we do not directly consider the role of star formation in consuming the unstable fraction of the gas. If the inflow rate is above a critical value, the resulting fragmentation would lead to some fraction of the inflowing gas collapsing to form stars rather than feeding the black hole accretion \citep{Lodato2006, Devecchi2010}. This has the potential then to limit the inflow onto the gas reservoir which feeds the black hole. The growth of the black hole is therefore potentially over-estimated. 

An alternative to viscosity driven inflow is to model the torque exerted on a galactic disc from gravitational instabilities within the disc \citep{Hopkins2011, AnglesAlcazar2015}. \citet{Hopkins2011} compared both the viscosity and gravitational torque driven accretion models to galaxy major merger simulations and found that inflow rates calculated using the  viscosity model were under-estimated, while the gravitational torque model matched the simulations more closely. Calculating the inflow rate using the gravitational torque model would increase the accretion rate of the black hole however, this would depend more directly on the star formation calculation. The gravitational torque model successfully recreates the black-hole-galaxy scaling relations at lower redshift \citep{AnglesAlcazar2013, Dave2019}.

We model an isothermal disc. Decreasing the gas temperature either globally or locally would decrease the sound speed and the overall stability of the disc. As the model evolves, increasing gas densities and star formation will result in the introduction of more coolants to the gas, such as $\mathrm{H}^-$, $\mathrm{H}_2$, and metals. The gas can then cool efficiently to $T_{\rm gas}\sim 100$ K, leading to a decrease in the disc stability. The inflow rate goes as $\dot{M}_{\rm inf} \propto c_{\rm s}^2$ (see Equation~\ref{mdotinf}) and as such the decrease in temperature would decrease the model inflow rate. Note however, the isothermal disc assumption would likely be no longer valid following the formation of these additional coolants.


In this study we have focused on haloes at the atomic cooling limit ($T_{\rm vir}\sim 10^4$ K). This provides the model with an initial halo mass at the redshift where the seed is assumed to have formed. However, if the formation were to take place at earlier times, $z_{\rm i}>20$, a lower halo growth rate would be sufficient to reach the same halo mass at $z=6$ or, similarly, with the same average growth rate a higher final halo mass would be reached. With higher halo masses, the model would find more massive discs, and potentially higher inflow rates feeding black hole growth. However, increasing the formation redshift does more to feed star formation than black hole growth. For example, with a formation redshift of $z_{\rm i}=30$, an average halo growth rate ($\zeta=0.806$) and a maximum viscosity ($\alpha_\nu=0.4$), the black hole reaches only $1.93 \times 10^{8} \, \mathrm{M_\odot}$ by $z=6$. This is ignoring the upper limit on the possible halo mass at $z=6$ from the abundance of SMBHs, which is independent of the formation redshift. Furthermore, pushing the formation to higher redshifts would not be consistent with the DCBH formation mechanism. The high intensity LW-background necessary for DCBH formation is not available prior to the formation of the first luminous galaxies at around $z\sim20 - 15$ \citep{Springel2005, Lacey2011, Agarwal2014a}.

\section*{Acknowledgements}
\defcitealias{Astropy2013}{Astropy Collaboration, 2013}
DSE acknowledges the financial support of the Science and Technology Facilities Council through a studentship award. We would like to thank the anonymous referee for constructive comments which helped to improve this paper as well as Jose O{\~n}orbe, Tilman Hartwig, Bhaskar Agarwal and Marta Volonteri for their useful discussions and comments. This research made use of Astropy, a community-developed core Python package for Astronomy \citepalias{Astropy2013}.




\bibliographystyle{mnras}
\bibliography{library} 








\bsp	
\label{lastpage}
\end{document}